\documentclass[structabstract]{aa}  

\usepackage{graphicx}
\usepackage{txfonts}
\usepackage{epsfig}

\usepackage[usenames]{color}
\newcommand\pycasso{{\sc p}y{\sc casso}}         
\newcommand\starlight{{\sc starlight}}          	

\begin{document}

\title{Resolving galaxies in time and space: II:}
\subtitle{Uncertainties in the spectral synthesis of datacubes}

\authorrunning{Cid Fernandes et al.}
\titlerunning{Uncertainties in spectral fitting of datacubes}

\author{R. Cid Fernandes\inst{1,2},
R. M. Gonz\'alez Delgado\inst{2},
R. Garc\'{\i}a Benito\inst{2},           
E. P\'erez\inst{2},
A. L.\ de Amorim\inst{1,2}, 
S. F. S\'anchez\inst{2,3},
B. Husemann\inst{4},
J. Falc\'on Barroso\inst{5,6},
R. L\'opez-Fern\'andez\inst{2},
P.  S\'anchez-Bl\'azquez\inst{7},
N. Vale Asari\inst{1},
A. Vazdekis\inst{5,6},
C. J. Walcher\inst{4},
\and
D. Mast\inst{2,3}}

\institute{
Departamento de F\'{\i}sica, Universidade Federal de Santa Catarina, P.O. Box 476, 88040-900, Florian\'opolis, SC, Brazil\\
\email{cid@astro.ufsc.br}
\and
Instituto de Astrof\'{\i}sica de Andaluc\'{\i}a (CSIC), P.O. Box 3004, 18080 Granada, Spain
\and
Centro Astron\'omico Hispano Alem\'an, Calar Alto, (CSIC-MPG), C/Jes\'us Durb\'an Rem\'on 2-2, E-04004 Almer\'{\i}a, Spain
\and
Leibniz-Institut f\"{u}r Astrophysik Potsdam, innoFSPEC Potsdam, An der Sternwarte 16, 14482 Potsdam, Germany
\and
Instituto de Astrof\'{\i}sica de Canarias, V\'{\i}a Lactea s/n, E-38200 La Laguna, Tenerife, Spain
\and
Departamento de Astrof\'{\i}sica, Universidad de La Laguna, E-38205, Tenerife, Spain
\and
Departamento de F\'{\i}sica Te\'orica, Universidad Aut\'onoma de Madrid, 28049 Madrid, Spain
}

\date{Received April 15, 2013; accepted June 28, 2013}

 
\abstract
{}
{
In a companion paper we have presented  many products derived from the application of the spectral synthesis code \starlight\ to datacubes from the CALIFA survey, including 2D maps of stellar population properties (such as mean ages, mass, extinction) and 1D averages in the temporal and spatial dimensions.
Our goal here is to assess the uncertainties in these products.
}
{
Uncertainties associated to noise and spectral shape calibration errors in the data and to the synthesis method are investigated by means of a suite of simulations, perturbing spectra and processing them through our analysis pipelines. The simulations use 1638 CALIFA spectra for NGC 2916, with perturbations amplitudes gauged in terms of the expected errors. A separate study was conducted to assess uncertainties related to the choice of evolutionary synthesis models, the key ingredient in the translation of spectroscopic information to stellar population properties. We compare results obtained with three different sets of models: the traditional Bruzual \& Charlot models, a preliminary update of them, and a combination of spectra derived from the Granada and MILES models. About $10^5$ spectra from over 100 CALIFA galaxies are used in this comparison. 
}
{
Noise and shape-related errors at the level expected for CALIFA propagate to uncertainties of 0.10--0.15 dex in stellar masses, mean ages and metallicities. Uncertainties in $A_V$ increase from 0.06 mag in the case of random noise to 0.16 mag for spectral shape errors. Higher order products such as star formation histories  are more uncertain than global properties, but still relatively stable. Due to the large number statistics of datacubes, spatial averaging reduces uncertainties while preserving information on the history and structure of stellar populations. Radial profiles of global properties, as well as star formation histories averaged over different regions are much more stable than those obtained for individual spaxels.
Uncertainties related to the choice of base models are larger than those associated with data and method. Except for metallicities, which come out very different when fits are performed with Bruzual \& Charlot models, differences in mean age, mass and metallicity are of the order of 0.15 to 0.25 dex, and 0.1 mag 	for $A_V$. 
Spectral residuals are of order of 1\% on average, but with systematic features of up to 4\% amplitude. The origin of these features, most of which are present in both in CALIFA and SDSS spectra, is discussed.
}
{}

\maketitle

\section{Introduction}
\label{sec:Intro}

Large optical spectroscopic surveys have dominated the panorama of extra-galactic astrophysics over the last decade. Stellar population analysis of these data has produced  information on stellar masses, mean stellar ages, abundances, dust content and diagnostics of their star formation histories (SFHs), whose analysis per se and in relation to other properties (say, morphology or  environment) has advanced the understanding of the physics of galaxies and their evolution through cosmic time. Yet, all this information is derived from spatially unresolved spectra, and thus reflect global averages over properties which are known to vary accross the face of galaxies. Integral field spectroscopy (IFS) is a natural next step to go from global properties to the internal physics of galaxies. Dissecting the different structural components  should lead to a much more complete picture of galaxies, as well as to a clearer understanding of the results already obtained in non-spatially resolved surveys. IFS has been around for over two decades, but has only recently been promoted to the scale of modern galaxy surveys. CALIFA, the Calar Alto Integral Field Area survey (S\'anchez et al.\ 2012; Husemann et al.\ 2013), is a pioneer in this blooming field.

To harvest the full potential of IFS data one needs to adapt techniques developed for spatially unresolved galaxy spectra, folding in the information on the spatial structure. In a companion article (Cid Fernandes et al.\ 2013, hereafter Paper I) we have presented our method to extract stellar population properties out of CALIFA datacubes by means of the \starlight\ spectral fitting software (Cid Fernandes et al.\ 2005). The packages developed to pre-process and partition the reduced datacubes ({\sc qbick}) and to post-process the results of the spectral synthesis (\pycasso\footnote{Python CALIFA Starlight Synthesis Organizer}) were described in detail. The products of this work flow were examplified using data for the nearby spiral NGC 2916 (CALIFA 277). Diagnostics such as 2D maps of physical properties (mean ages and metallicities, mass surface densities, star formation rates, etc.), 1D averages in the temporal and spatial dimensions, projections of the stellar light and mass growth onto radius-age diagrams, and cuts through the $(x,y,{\rm age})$ cubes were presented to illustrate the potential of the combination of IFS data with spectral synthesis as a tool to study galaxy evolution in time and space simultaneously.

Uncertainties in this whole process stem from {\em (i)} noise and calibration of the data, {\em (ii)} limitations of our spectral synthesis method, and {\em (iii)} evolutionary synthesis models, the key ingredient in the mapping of observed spectra to astrophysical information. These different sources of uncertainty, each one comprising a large field of work by itself, intertwine and propagate in non trivial ways through our pipelines. 
Uncertainties in spectral synthesis products have been previously explored in the literature, like the noise-effect simulations in Cid Fernandes et al.\ (2005), the model-impact study of Panter el al.\ (2007), and the suite of tests presented by Tojeiro et al.\ (2007), but never in the context of IFS.

This article complements Paper I by addressing the effects of these uncertainties. Our central goal is to provide quantitative and qualitative guidelines to allow an assessment of the robustness of results produced with the methods and tools explained in Paper I, which are being used in a series of publications by our collaboration. P\'erez et al.\ (2013), for instance, analysed the spatially resolved mass assembly history of 105 CALIFA galaxies, finding clear evidence for an inside-out growth for massive galaxies, and signs of outside-in growth for the less massive ones (see   Gonz\'alez Delgado et al.\ 2012a and 2012b for other results). Despite the focus on \starlight\ and CALIFA, the results reported here are also useful in the broader context of stellar population analysis of IFS data; for instance, experimental datacubes recently obtained by MaNGA run through our {\sc qbick} and \pycasso\ pipelines without any modification.

The first part of this paper (\S\ref{sec:SimulationsDescription} and \S\ref{sec:SimulationsResults}) presents extensive simulations which explore effects of uncertainties related to the data and spectral synthesis method. Random noise and shape-changing perturbations are added to both observed and synthetic spectra of all zones of the same galaxy analysed in Paper I. These simulations are then used to evaluate uncertainties in global properties (mean ages, masses, etc.) and higher order products such as SFHs. This is done both at the level of individual spectra and for spatial regions cut out from the  datacube. 

The second part (\S\ref{sec:SSPtests}) is dedicated to experiments carried out to evaluate how our results are affected by different choices of evolutionary synthesis models for the spectra of simple stellar populations (SSP). This study uses $\sim$ 100 thousand spectra from 107 galaxies observed by CALIFA, all of which were processed with three sets of SSP models used in the current literature. An inspection of the quality of the spectral fits and a comparison with results obtained for SDSS galaxies is also presented. 

\S\ref{sec:Conclusions} summarizes our main results. Throughout this paper the notation and definitions established in Paper I are followed.

\section{Uncertainties associated to the data and the method: Description of the simulations}
\label{sec:SimulationsDescription}

A drawback of \starlight\ is that it provides no error estimate on its output. The most straightforward (albeit computationaly expensive) way to do so is to run the code several times, perturbing the input data according to some realistic prescription of the errors involved. This section describes a set of extensive simulations taylored to match the characteristics of CALIFA data. The simulations are designed to address the effects of 

\begin{enumerate}
\item random noise, 
\item shape-related calibration uncertainties, and 
\item multiplicity of solutions
\end{enumerate}

\noindent upon the physical properties derived with \starlight. Throughout these experiments we keep the set of SSPs fixed at the same one used in Paper I and denoted ``base {\it GM}'' in \S\ref{sec:SSPtests}, where the effects of this choice are investigated. This base comprises $N_\star = 156$ elements with 39 ages between $t = 1$ Myr and 14 Gyr, and 4 metallicities from $Z = 0.2$ to 1.5 solar. The simulations are  based on the data and \starlight\ fits of 1638 zone spectra\footnote{1527 of the spectra come from single spaxels, while the remaining 111 ones correspond to Voronoi zones comprising typically 10 spaxels.} of CALIFA 277 studied in Paper I. Given the variety of properties found accross this galaxy and the fact that all other datacubes in CALIFA were processed in the same way, the uncertainty estimates presented here are representative of CALIFA data in general.

Each spectrum was perturbed 10 times for each of the R, C and E error prescriptions discussed below, and refitted. This data set is then used to evaluate uncertainties in the derived properties, both at the level of individual zones and for spatially averaged properties like those shown in Figs.\ 10 and 11 of Paper I. Both observed and synthetic spectra are perturbed in the simulations. 
The former represent real-life conditions, while the latter is essentially a theoretical exercise which explores the sensitivity of the \starlight\ results to different kinds of perturbations.

\subsection{Types of errors and simulations}
\label{sec:ErrorsAndSimulations}

The first type of error considered is that related to the variations on the derived properties induced by random fluctuations of the observed fluxes within the noise level. To study this, each zone spectrum was perturbed 10 times with gaussian noise with a $\lambda$-dependent amplitude given by the error spectrum $\epsilon_\lambda$ (hereafter OR1 runs, with ``O'' standing for observed, ``R'' for random, and 1 for the amplitude of the perturbation). Experiments were also conducted with pertubations of 2 and $3 \times \epsilon_\lambda$ (OR2 and OR3 runs, respectively) to emulate what would be obtained under considerably worse conditions than those in CALIFA.\footnote{Note that these simulations add noise to observed spectra, which already contains it. Hence, OR1 spectra, which are nominaly  meant to represent variations at the level of $1 \epsilon_\lambda$, effectively have errors of $\sqrt{2} \epsilon_\lambda$, ie., 41\% larger than intended. We will nevertheless pretend these runs correspond to $1 \sigma$ perturbations, which gives a safety margin in our analysis. Similarlty, OR2 and OR3 runs have errors 2.23 and 3.16 times larger than the original ones, instead of factors of 2 and 3, respectively.} 

A second set of simulations was carried to address the issue of overall continuum shape calibration. \starlight\ relies on flux calibrated input spectra, so it is relevant to access how much its solutions vary for spectra whose shape vary within expected uncertainty ranges. Husseman et al.\ (2013) find that CALIFA synthetic $g - r$ colors typically differ  by 0.05 mag from those obtained by SDSS photometry. We use this value to gauge the effects of continumm calibration uncertainty in each individual specrum of CALIFA 277. For simplicity, but with no loss of generality, these simulations keep  the flux  at $\lambda_r = 6231$ \AA\ stable, such that the 0.05 mag variation in $g - r$ translates  to a 0.02 dex uncertainty in the flux at $\lambda_g = 4770$ \AA. To emulate the effects of such variations we have produced 10 versions of each spectrum by adding $A_g \times (\lambda_r - \lambda) / (\lambda_r - \lambda_g)$ to the original $\log F_\lambda$, with $A_g = \Delta \log F_{\lambda_g}$ sampled from a gaussian distribution of zero mean and dispersion $\sigma(\log F_{\lambda4470}) = 0.02$  (OC002 runs, where ``C'' stands for color). This produces both redder and bluer versions of the original spectrum. Experiments were also carried out for $\sigma(\log F_{\lambda4470}) = 0.01$, and 0.04 dex (OC001 and OC004, respectively). We take OC002 runs as default for CALIFA, even though the spaxel-to-spaxel variations in the continuum calibration for a single galaxy are probably smaller than the $g - r$ dispersion quoted by Husemann et al.\ (2013), which was derived from the 100 galaxies in the CALIFA DR1 sample.

Finally, fits were also conducted for no perturbation at all (OE0 runs), the motivation for which is to quantify a qualitatively different source of uncertainty, embedded in all other simulations. \starlight\ provides a single best fit set of parameters, out of typically many millions tried out during its likelyhood-guided sampling of the parameter space. This single solution, however, is not meant to be mathematically unique. In fact, given the pseudo random nature of its Markov chains, \starlight\ solutions depend on the input seed for the random number generator. For large samples, and each CALIFA datacube is a large sample in itself, these variations have no relevant effect on the overall results of the analysis. For individual spectra, however, the multiplicity of solutions leads to uncertainties in the derived properties. The OE0 runs quantify these uncertainties,
which ultimately reflect an intrinsic limitation of our spectral synthesis method.

\subsection{Simulations based on synthetic spectra}
\label{sec:S_simulations}


The OE0, OR* and OC* runs work upon the original observed spectra (hence the prefix ``O''). A paralell set of simulations was carried out applying these same perturbation recipes to the synthetic spectra obtained in our \starlight\ fits of CALIFA 277. We label these runs as SE0, SR1, SR2, SR3, SC001, SC002 and SC004 (``S'' for synthetic). 

The S runs use exactly the same sampling, masks, flags and $\epsilon_\lambda$ errors as in the O runs, thus following the pattern of CALIFA data in a realistic fashion. At the same time, these are idealistic simulations in the sense that the perturbations act upon ``perfect'' input spectra, instead of observed ones. This theoretical exercise therefore overlooks inadequacies in the SSP models 
(including issues like potential flux calibration problems in the libraries, limited coverage of the stellar parameter space, missing or incorrectly modeled evolutionary phases) and other assumptions involved in the modelling (namely, extinction and kinematics). 

The purpose of these runs is to evaluate to which extent \starlight\  recovers known input properties in the presence of perturbations. This kind of input versus output comparison is more commonly carried out feeding the spectral fitting code with data generated with ad hoc parametric descriptions of the SFH (eg, the so called ``tau models''), while our theroretical SFHs are inspired in those derived from actual observations. A valid criticism of these simulations is that the same base is used to generate and fit spectra. The alternative would be to generate models with some other set of $t's$ and $Z$', but since our base is so large (39 ages and all 4 available metallicities), this would make very little difference in practice.

\subsection{Notation}
\label{sec:Notation4Sims}

\begin{table}
\begin{centering}
\begin{tabular}{lrrrr}
\multicolumn{5}{c}{Notation for the simulations} \\ \hline
Run & original & type of      & perturbation & comments \\
    & spectra  & perturbation & amplitude    &          \\
(1) & (2)      & (3)          & (4)          & (5)      \\
\hline
OR1   & observed  & noise        & $1 \times \epsilon_\lambda$ & gaussian \\
OR2   & observed  & noise        & $2 \times \epsilon_\lambda$ & gaussian \\
OR3   & observed  & noise        & $3 \times \epsilon_\lambda$ & gaussian \\
OC001 & observed  & color        & 0.01 dex                    & at $\lambda_g$ \\
OC002 & observed  & color        & 0.02 dex                    & at $\lambda_g$ \\
OC004 & observed  & color        & 0.04 dex                    & at $\lambda_g$ \\
OE0   & observed  & none         & 0                           &  \\
SR1   & synthetic & noise        & $1 \times \epsilon_\lambda$ & gaussian \\
SR2   & synthetic & noise        & $2 \times \epsilon_\lambda$ & gaussian \\
SR3   & synthetic & noise        & $3 \times \epsilon_\lambda$ & gaussian \\
SC001 & synthetic & color        & 0.01 dex                    & at $\lambda_g$  \\
SC002 & synthetic & color        & 0.02 dex                    & at $\lambda_g$  \\
SC004 & synthetic & color        & 0.04 dex                    & at $\lambda_g$  \\
SE0   & synthetic & none         & 0                           &  \\
\hline
\end{tabular}
\end{centering}
\caption{(1) Nomenclature adopted to denote the different simulations. (2) The nature (observed versus synthetic) of the spectra which are perturbed. (3) Type of perturbation. (4) Amplitude of the perturbations. For C runs, we list the amplitude of the variations at $\lambda_g$. 
Each of the 1638 spectra from CALIFA 277 are perturbed 10 times according to each of these 14 recipes.
}
\label{tab:Notation4Sims}
\end{table}

For clarity and quick reference, we summarize the notation and intended purpose of
the simulations described above. 

R runs (OR1, OR2, OR3, SR1, SR2 and SR3) explore the effect of noise by adding gaussian perturbations scaled to have amplitudes 1, 2 or 3 times the nominal error $\epsilon_\lambda$.  C runs perturb the input spectra in their global shape by multiplying the original $F_\lambda$ by a power-law in $\lambda$, scaled such that the flux at $\lambda_r$ remains fixed while that at $\lambda_g$ differs by (on average over all realizations) 0.01 (OC001 and SC001), 0.02 (OC002 and SC002) and 0.04 (OC004 and SC004) dex with respect to the input flux. (In a $\log F_\lambda \times \lambda$ plot, these shape-pertubations correspond to straight lines anchored at $\lambda_r$ and different slopes.) These runs are meant to emulate the effect of a flux calibration uncertainty in the data. E0 runs do nothing to the input spectra. They only differ in the random number seed feed to \starlight. The variance among these fits reflects the multiplicity of SSP combinations leading to a similar total spectrum.

In all cases, a prefix ``O'' indicates that the perturbations act upon the observed spectra, while an ``S'' prefix means that synthetic spectra are perturbed. Table \ref{tab:Notation4Sims} summarizes the simulation-related nomenclature used thoughout the paper.

\section{Results of the simulations}
\label{sec:SimulationsResults}

The simulations described in \S\ref{sec:SimulationsDescription}
provide the material to evaluate uncertainties associated with the data and spectral synthesis method. We split the presentation of this material into an assessement of the errors in global properties (\S\ref{sec:ErrorsGlobalProperties}), radial profiles (\S\ref{sec:ErrorsRadialProfiles}), and SFHs (\S\ref{sec:ErrorsSFHs}). We conclude this part of our investigation with a summary of the results of  the simulations (\S\ref{sec:ErrorsConclusion}).

\subsection{Uncertainties in global properties}
\label{sec:ErrorsGlobalProperties}

\begin{figure*}
\includegraphics[width=1.0\textwidth]{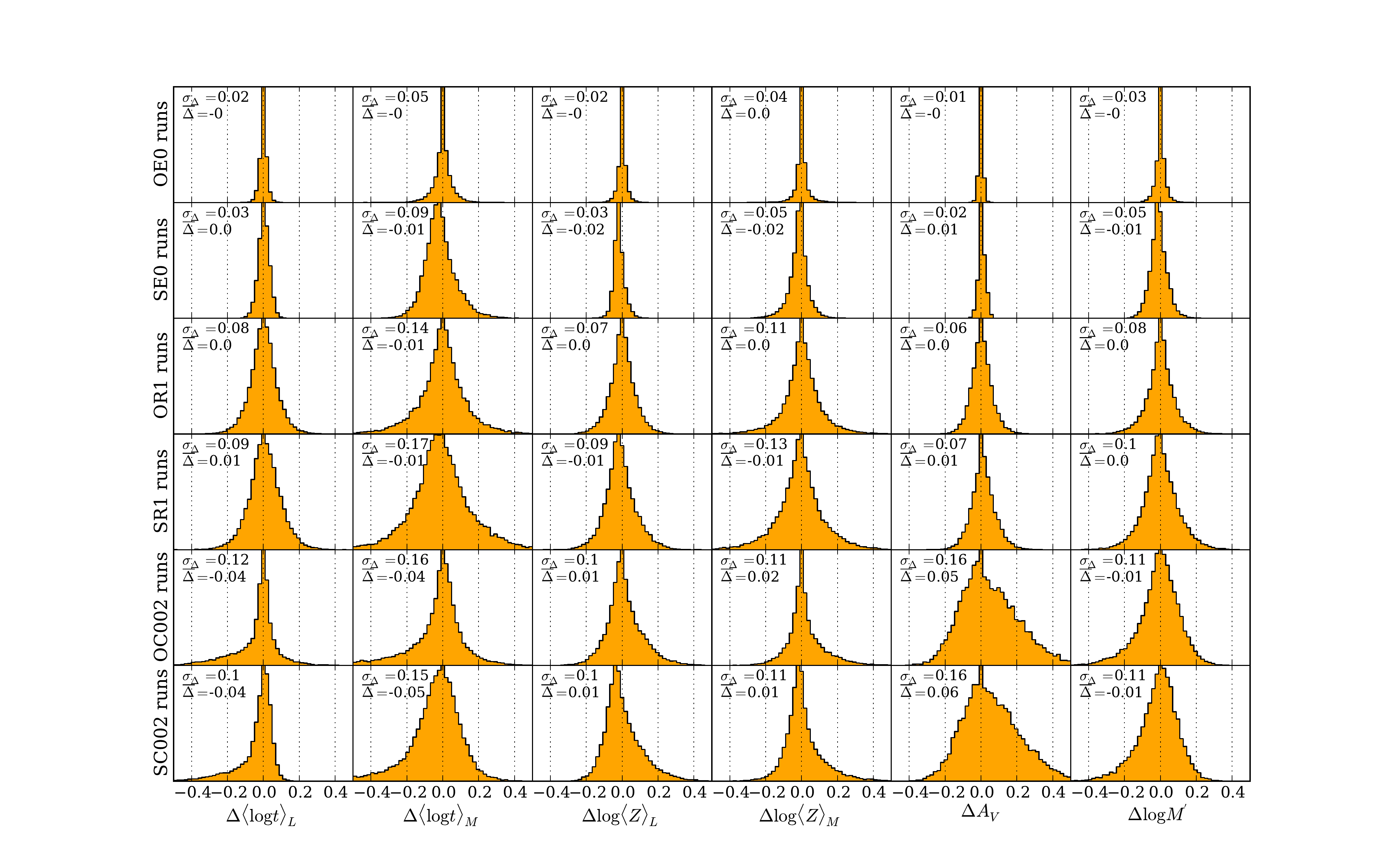}
\caption{Histograms of $\Delta$ ($\equiv$ Simulated minus Original value) for the main scalar properties derived from the spectral synthesis: $\langle \log t\rangle_L$, $\langle \log t\rangle_M$, $\log \langle Z\rangle_L$, $\log \langle Z\rangle_M$, $A_V$ and the initial stellar mass formed ($\log M^\prime$). Each row corresponds to one set of simulations, alternating S and O runs (as labeled in the y-axis). Mean $\Delta$'s and their $\sigma$'s are labeled in each panel. Except for the $A_V$ column, each histogram is based on 16380 values obtained from the 1638 zones in CALIFA 277 and 10 Monte Carlo realizations of the perturbation. For clarity, the $A_V$ histograms were drawn excluding the many $\Delta A_V = 0$ cases, which occur due to the $A_V \ge 0$ physical limit imposed on the spectral fits.
Noise-free simulations (OE0 and SE0), which trace intrinsic degeneracies in spectral fits of composite stellar populations, are show in the top two rows.
Random-noise simulations (SR1 and OR1, central rows) map the uncertainties due to statistical fluctuations of the observed $F_\lambda$ fluxes.
C002-simulations, where the spectral-shape is changed to emulate calibration uncertainties, are shown on the bottom panels.}
\label{fig:DiffHists} 
\end{figure*}

\begin{figure*}
\includegraphics[width=1.0\textwidth]{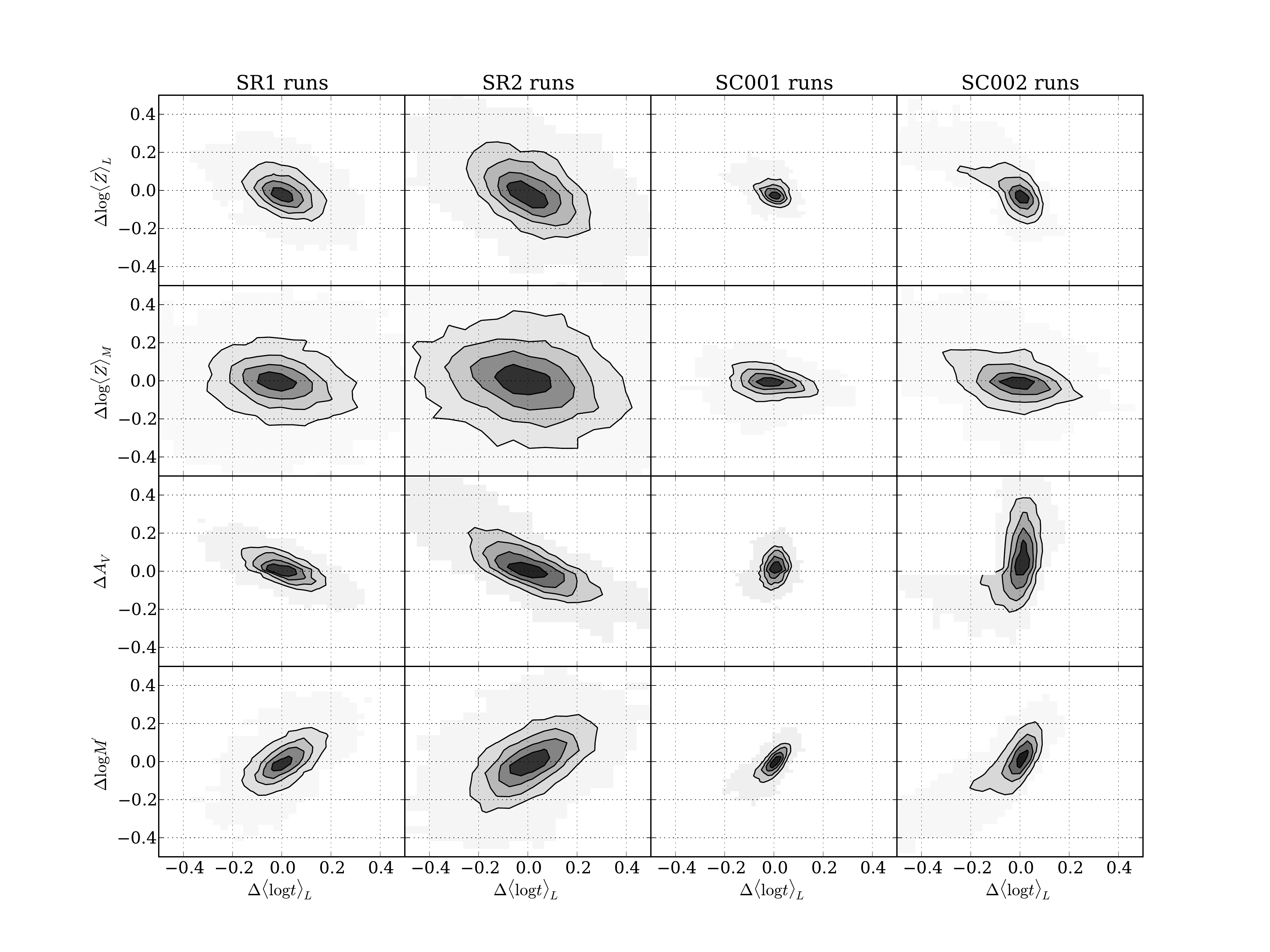}
\caption{Illustration of the correlated $\Delta$ variations of ages, metallicites, extincion, and masses. Each column corresponds to one set of simulations. 
Iso-density contours mark 20, 40, 60 and 80\% of enclosed points.
}
\label{fig:dXdY} 
\end{figure*}

\begin{figure*}
\includegraphics[width=1.0\textwidth]{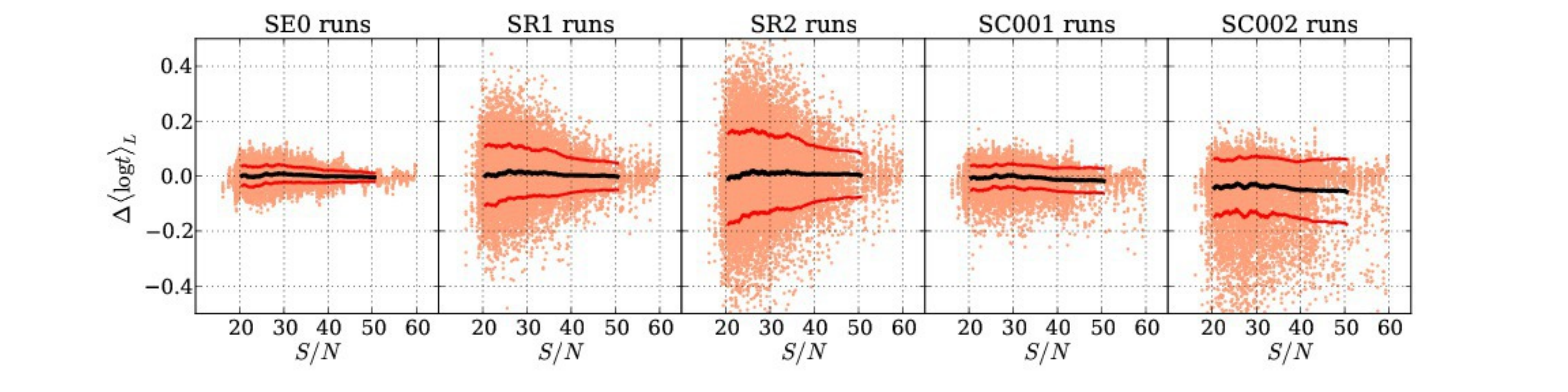}
\caption{$\Delta \langle \log t\rangle_L$ variations for all 1638 zones $\times$  10 realizations plotted against the $S/N$ of the corresponding original (observed) spectra. Different panels correspond to different sets of simulations, as labeled. The lines are running mean of $\Delta \langle \log t\rangle_L$ and the corresponding $\pm 1 \sigma$ intervals.}
\label{fig:SN_sims} 
\end{figure*}

\begin{table*}
\begin{centering}
\begin{tabular}{lrrrrrrr}
\multicolumn{8}{c}{Summary of simulations: $\overline{\Delta} \pm \sigma_\Delta$} \\ \hline
%
 Property   &  OE0  &  OR1  &  OR2  &  OR3  &  OC001  &  OC002  &  OC004  \\
\hline
$\langle \log t \rangle_L$  &   -0.00   $\pm$ {\bf    0.02  }   &   0.00   $\pm$ {\bf    0.08  }   &   0.01   $\pm$ {\bf    0.13  }   &   0.01   $\pm$ {\bf    0.16  }   &   -0.00   $\pm$ {\bf    0.05  }   &   -0.04   $\pm$ {\bf    0.12  }   &   -0.09   $\pm$ {\bf    0.20  }   \\
$\langle \log t \rangle_M$  &   -0.00   $\pm$ {\bf    0.05  }   &   -0.01   $\pm$ {\bf    0.14  }   &   -0.02   $\pm$ {\bf    0.21  }   &   -0.03   $\pm$ {\bf    0.26  }   &   -0.00   $\pm$ {\bf    0.08  }   &   -0.04   $\pm$ {\bf    0.16  }   &   -0.10   $\pm$ {\bf    0.28  }   \\
$\log \langle Z \rangle_L$  &   -0.00   $\pm$ {\bf    0.02  }   &   0.00   $\pm$ {\bf    0.07  }   &   -0.00   $\pm$ {\bf    0.12  }   &   -0.00   $\pm$ {\bf    0.16  }   &   0.00   $\pm$ {\bf    0.05  }   &   0.01   $\pm$ {\bf    0.10  }   &   0.03   $\pm$ {\bf    0.16  }   \\
$\log \langle Z \rangle_M$  &   0.00   $\pm$ {\bf    0.04  }   &   0.00   $\pm$ {\bf    0.11  }   &   -0.00   $\pm$ {\bf    0.18  }   &   0.00   $\pm$ {\bf    0.22  }   &   0.00   $\pm$ {\bf    0.05  }   &   0.02   $\pm$ {\bf    0.11  }   &   0.05   $\pm$ {\bf    0.18  }   \\
$A_V                     $  &   -0.00   $\pm$ {\bf    0.01  }   &   0.00   $\pm$ {\bf    0.06  }   &   0.01   $\pm$ {\bf    0.10  }   &   0.02   $\pm$ {\bf    0.14  }   &   0.01   $\pm$ {\bf    0.06  }   &   0.05   $\pm$ {\bf    0.16  }   &   0.14   $\pm$ {\bf    0.29  }   \\
$\log M^\prime           $  &   -0.00   $\pm$ {\bf    0.03  }   &   0.00   $\pm$ {\bf    0.08  }   &   -0.00   $\pm$ {\bf    0.13  }   &   0.00   $\pm$ {\bf    0.15  }   &   -0.00   $\pm$ {\bf    0.05  }   &   -0.01   $\pm$ {\bf    0.11  }   &   -0.02   $\pm$ {\bf    0.21  }   \\
\hline
%
%
 Property   &  SE0  &  SR1  &  SR2  &  SR3  &  SC001  &  SC002  &  SC004  \\
\hline
$\langle \log t \rangle_L$  &   0.00   $\pm$ {\bf    0.03  }   &   0.01   $\pm$ {\bf    0.09  }   &   0.01   $\pm$ {\bf    0.14  }   &   0.01   $\pm$ {\bf    0.18  }   &   -0.01   $\pm$ {\bf    0.04  }   &   -0.04   $\pm$ {\bf    0.10  }   &   -0.10   $\pm$ {\bf    0.19  }   \\
$\langle \log t \rangle_M$  &   -0.01   $\pm$ {\bf    0.09  }   &   -0.01   $\pm$ {\bf    0.17  }   &   -0.02   $\pm$ {\bf    0.23  }   &   -0.03   $\pm$ {\bf    0.27  }   &   -0.02   $\pm$ {\bf    0.10  }   &   -0.05   $\pm$ {\bf    0.15  }   &   -0.11   $\pm$ {\bf    0.26  }   \\
$\log \langle Z \rangle_L$  &   -0.02   $\pm$ {\bf    0.03  }   &   -0.01   $\pm$ {\bf    0.09  }   &   -0.01   $\pm$ {\bf    0.14  }   &   -0.00   $\pm$ {\bf    0.18  }   &   -0.01   $\pm$ {\bf    0.05  }   &   0.01   $\pm$ {\bf    0.10  }   &   0.02   $\pm$ {\bf    0.17  }   \\
$\log \langle Z \rangle_M$  &   -0.02   $\pm$ {\bf    0.05  }   &   -0.01   $\pm$ {\bf    0.13  }   &   -0.00   $\pm$ {\bf    0.20  }   &   0.01   $\pm$ {\bf    0.24  }   &   -0.01   $\pm$ {\bf    0.06  }   &   0.01   $\pm$ {\bf    0.11  }   &   0.04   $\pm$ {\bf    0.19  }   \\
$A_V                     $  &   0.01   $\pm$ {\bf    0.02  }   &   0.01   $\pm$ {\bf    0.07  }   &   0.02   $\pm$ {\bf    0.11  }   &   0.03   $\pm$ {\bf    0.15  }   &   0.02   $\pm$ {\bf    0.06  }   &   0.06   $\pm$ {\bf    0.16  }   &   0.15   $\pm$ {\bf    0.30  }   \\
$\log M^\prime           $  &   -0.01   $\pm$ {\bf    0.05  }   &   0.00   $\pm$ {\bf    0.10  }   &   0.00   $\pm$ {\bf    0.14  }   &   0.00   $\pm$ {\bf    0.16  }   &   -0.01   $\pm$ {\bf    0.06  }   &   -0.01   $\pm$ {\bf    0.11  }   &   -0.03   $\pm$ {\bf    0.20  }   \\
\hline
\end{tabular}
\end{centering}
\caption{Statistics of the simulations. For each global property the table lists the mean simulated minus original difference ($\overline{\Delta}$) and its standard deviation ($\sigma_\Delta$). The latter, printed in bold, indicates the level of uncertaintites under different circunstances. Each column denotes one set of simulations. The upper half of the table shows results of simulations based on variations upon the original observed spectrum (O runs), whereas the bottom half corresponds to runs where the synthetic spectra were perturbed (S runs) and used as input.}
\label{tab:Stats_SimsK0277}
\end{table*}

The main scalar properties derived from the spectral synthesis are mean ages and metallicities (weighted by light or mass), extinction and stellar masses\footnote{ Throughout this section we work with the initial stellar mass ($M^\prime$), uncorrected for the mass expelled from stars back to the interstellar medium. Mass weighted ages and metallicities, however, do account for the returned mass.}. These 1$^{st}$ order products of the analysis convey enough information to draw a quite detailed sketch of the 2D distribution of stellar population properties in a galaxy. Before incorporating the spatial arrangement of the spectra, however, it is useful to assess uncertainties in the individual spectra, as if they came from different galaxies.

Fig.\ \ref{fig:DiffHists} shows histograms of the $\Delta \equiv$ Simulation $-$ Original value (ie., that obtained from the unperturbed spectrum) of each of these quantities, and for six different sets of simulations: E0, R1 and C002 runs, in both O and S versions. Statistics for all runs are listed in Table \ref{tab:Stats_SimsK0277}.

\subsubsection{E runs: Noise-free simulations}
\label{sec:Eruns}

The OE0 and SE0 panels show that variations due to the random number seed feed to \starlight\ have a minute effect on  the derived properties, with no bias ($\overline\Delta \sim 0$) and one $\sigma$ uncertainties all in the second decimal place. 

Intriguingly, SE0 runs have somewhat larger dispersions than OE0 runs. The mathematical reason for this behaviour is subtle, but simple to grasp with the concepts developed by Pelat (1997, 1998). By construction, SE0 spectra lie within the Synthetic Domain (SD), ie, the space generated by convex combinations of the base spectra\footnote{In the case of \starlight\ and other full spectral fitting codes this space is further enlarged by the extinction and kinematical dimensions, but we leave this aside to simplify the argument.}. A point inside this space can be exactly fitted by at least one set of parameters. Even if the solution is mathematically unique, a whole range of other solutions produce nearly identical spectra.  Because of the limitations of evolutionary synthesis models, observed spectra cannot possibly be fitted exactly even in the ideal limit of perfect calibration and $S/N = \infty$, so OE0 spectra are inevitably {\em outside} the SD. Fits of OE0 spectra therefore have access to a smaller sub-space of acceptable solutions than fits of SE0 spectra, resulting in the smaller dispersions observed in Fig.\ \ref{fig:DiffHists}. This same interpretation applies to other runs. Even when the synthetic spectra are perturbed (say, as in the SR1 simulations), they remain closer to the SD than the corresponding O spectra, thus reaching a larger region of the solutions space.

\subsubsection{R runs: Random noise simulations}
\label{sec:Rruns}

Random noise at the nominal level (OR1 and SR1 runs, middle rows in Fig.\ \ref{fig:DiffHists}) lead to $\Delta$-distributions with $\sigma$ around 0.1 dex in ages, metallicities and masses, and 0.05 mag in $A_V$. As expected (and also seen in the E0 runs), mass-weighted ages and metallicities have broader $\Delta$ distributions than their luminosity weighted counterparts. In any case, these dispersions are all relatively small. Indeed, they are smaller than those induced by the use of different SSP models (\S\ref{sec:SSPtests}, Fig.\ \ref{fig:SSPsDiffHists}).

Table \ref{tab:Stats_SimsK0277} reports results for simulations with noise levels 2 and 3 times worse than $\epsilon_\lambda$. For the OR2 runs, for instance the  the uncertainties in the properties ploted in Fig.\ \ref{fig:DiffHists} increase by roughly a factor of 2 with respect to those found in the OR1 simulations. The $\Delta$ distributions (not plotted) remain symmetric, with no significant biases. Recall, however, that there is no reason to believe that the $\epsilon_\lambda$ values produced by the CALIFA reduction pipeline are underestimated. On the contrary, Husemann et al.\ (2013) show that, if anything, they are slightly overestimated.

\subsubsection{C runs: Shape perturbations}
\label{sec:Cruns}

The bottom panels in Fig.\ \ref{fig:DiffHists} show the results for simulations where the continuum shape was perturbed to produce (on average) 0.02 dex variations in the flux at $\lambda_g$ for a fixed flux at $\lambda_r$. The one $\sigma$ variations in ages, metallicities and masses for OC002 runs are larger than those obtained with the OR1 runs, but not as large as in OR2 (Table \ref{tab:Stats_SimsK0277}). Similar conclusions apply to the corresponding S runs. Predictably, $A_V$ is more severely affected by these global shape perturbations, with $\sigma(\Delta A_V) \sim 2.5$ times larger than those in the OR1 runs. 

Unlike  in other simulations, the $\Delta$ distributions are skewed, particularly the ones for  $A_V$ and $\langle \log t \rangle_L$. This is a by-product of the $A_V \ge 0$ limit imposed upon the fits. 582 of the 1638 zones of CALIFA 277 were fitted with $A_V = 0$ (see Fig.\ 4 in Paper I). About half of the shape-perturbed versions of these spectra are bluer than the original ones. Since \starlight\ was not allowed to decrease $A_V$ to fit them, it gets stuck at $A_V = 0$, compensating the extra blueness with younger populations, producing the tail towards negative $\Delta \langle \log t \rangle_L$ seen in the bottom rows of Fig.\ \ref{fig:DiffHists}. This is also the reason why there are more instances of $\Delta A_V > 0$ than $< 0$, producing the positively skewed 
$\Delta A_V$ distributions.

Increasing the shape errors by a factor of 2 (C004 runs) leads to uncertainties roughly twice as large, biases of the order of $- 0.1$ dex in ages and $+ 0.15$ mag in $A_V$ (Table \ref{tab:Stats_SimsK0277}), as well as  larger skewness of the $\Delta$ distributions. Conversely, C001 runs result in uncertainties $\sim 2$ times smaller, with no bias nor much skewness.

\subsubsection{Covariances}
\label{sec:Covariances}

As usual in multivariate problems, variations in one property correlate with variations in others. In the context of spectral synthesis, such covariances are known as age-metallicity-extinction degeneracies: $A_V$ affects colors while $t$ and $Z$ affects both absorption lines and colors. Previous work (e.g., Cid Fernandes et al.\ 2005; S\'anchez-Bl\'azquez et al.\ 2011) showed that $\lambda$-by-$\lambda$ fits help reducing these degeneracies, but do not eliminate them. 

Fig.\ \ref{fig:dXdY} shows how the main scalar properies co-vary in our simulations. S runs are used in this plot---O runs produce similar plots, with slightly more compact contours. The first two columns (SR1 and SR2 runs) reveal the classical degeneracies of population synthesis, with mean ages increasing as mean metallicities decrease, and extinction variations also anti-correlated with variations in mean age. The bottom row shows how stellar masses increase as ages become older, and vice-versa.

In the R-runs, perturbations do not change the overall shape of the input spectrum. Hence, if a \starlight\ solution turns out somewhat older than the original one, its extinction and/or metallicity will decrease to compensate the redder colors and stronger absorption lines of the older stars. C-runs, on the other hand, do change colors---that is, in fact, exactly what they are designed to do. This changes the nature of ``degeneracies'' a bit. 

If the shape perturbation followed the same functional form of the extinction law, it would suffice to change $A_V$, leaving the population mixture essentially unchanged, varying $t$'s and $Z$'s at the level they would naturally vary without color perturbations (i.e, due to noise and the multiplicity of solutions). To first order, this is indeed what happens in the C-runs. Compare, for instance the (SR1, SR2) and (SC001, SC002) pairs of $\Delta A_V \times \Delta \langle \log t \rangle_L$ panels  in Fig.\ \ref{fig:dXdY}. In both pairs of plots the corresponding type of error changes by a factor of 2 in amplitude, yet one sees that the $\Delta A_V$ contours grow more from SC001 to SC002 than from SR1 to SR2. Furthermore, the $\Delta A_V$ contours are nearly vertically oriented in the C-runs, while in R-runs the anti-correlation with  $\Delta \langle \log t \rangle_L$ is clear (the age-extinction degeneracy). We thus confirm the expectation that, to first order, continuum shape uncertainties affect primarely the $A_V$ estimates.

There are still two ``second order'' features to explain about the $\Delta A_V \times \Delta \langle \log t \rangle_L$ C-runs maps in Fig.\ \ref{fig:dXdY}. First, the inner contours have a slightly positive slope, opposite to that expected from the age-extincion degeneracy (clearly seen in the SR1 and SR2 panels). This behaviour is due to the fact that, because of the different functional forms, our shape-perturbations (where $\Delta \log F_\lambda$ changes linearly with $\lambda$) cannot be fully absorbed by the reddening curve used in the fits. Since changing $A_V$ alone is not quite enough to produce optimal spectral fits, \starlight\ looks for other ways of making the synthetic spectrum redder or bluer (depending on the sign of the perturbation), and skewing the age mixture is the most effective one. In any case, the change in mean age is small: $\overline{\Delta}\langle \log t \rangle_L = -0.04$ for SC002 runs (Table \ref{tab:Stats_SimsK0277}). Secondly, for SC002 runs, the outer contours (containing about 20\% of the points) are limited to $\Delta A_V \leq 0$ whenever simulated mean ages are more than 0.1 dex younger than the original ones ($\Delta \langle \log t \rangle_L < -0.1$). Spectra along this odd looking ridge  were originally fitted with $A_V = 0$, and then perturbed to bluer colors. Since our fits do not allow $A_V < 0$, \starlight\ fitted them with the same $A_V$ (hence $\Delta A_V = 0$), but younger ages, leading to the pattern observed in this panel.

\subsubsection{Checking the effects of imposing $A_V \geq 0$}
\label{sec:AV0runs}

As seen just above, the effects of a hypothetical systematic redness associated to spectroscopic calibration issues are almost entirely absorbed biasing  $A_V$ upwards, while an excessive blueness of the same nature stumbles upon the $A_V \ge 0$ constraint, leading to biases in the age mixture. An artificial way of fixing this assimetry is to allow for $A_V < 0$ in the spectral fits. A trivial statistical justification to allow for this unphysical possibility is that, when $A_V$ is truly $= 0$, unbiased estimates should oscilate around 0, including both negative and positive values. We also note that completely independent studies (different methods, different data) also  allow for or deduce negative extinctions (e.g.\ Kauffmann et al.\ 2003, their figure 11).

An extra set of OE0-like simulations, but imposing $A_V > -1$ was run to investigate this issue. In practice what happens is that only zones which were originally fitted with $A_V \sim 0$ are systematicaly refitted with negative $A_V$. The 591 zones which come out with $A_V < 0$ in these new fits include all the 582  for which we had originally found $A_V = 0$, and the remaining 9 had $A_V$ very close to 0. The mean new $A_V$ of these 591 zones was $-0.12$ mag, with a standard deviation of 0.08 mag. The minimum value reached was $A_V = -0.4$ mag, but in only 16\% of the cases $A_V$ went below $-0.2$ mag. The variations in other global properties (for these same 591 zones) were 
$\overline{\Delta} \langle \log t \rangle_L = + 0.12$, 
$\overline{\Delta} \langle \log t \rangle_M = + 0.12$, 
$\overline{\Delta} \log \langle Z \rangle_L = - 0.07$,
$\overline{\Delta} \log \langle Z \rangle_M = - 0.04$,
and $\overline{\Delta} \log M^\prime  = + 0.03$ dex. 
Mean ages are the most affected property, but to an acceptably small degree, with biases of the same order as the one-$\sigma$ variations produced by random and shape-related uncertainties at the nominal level (ie, R1 and C002 simulations). 

Zones originally fitted with $A_V > 0$ did not show systematic variations in their global properties in these extra runs.

\subsubsection{The effect of $S/N$}
\label{sec:EffectsOfSN}

The $\Delta$ histograms in Fig.\ \ref{fig:DiffHists} and the statistics in Table \ref{tab:Stats_SimsK0277} represent the uncertainties of the  CALIFA 277 data set as a whole, including zones of different $S/N$. The general expectation is that smaller $S/N$ leads to larger uncertainties, such that the tails of the $\Delta$ distributions in Fig.\ \ref{fig:DiffHists} are dominated by the noisier spectra, while the cores (small $\Delta$ values) correspond to higher $S/N$.

To verify this, Fig.\ \ref{fig:SN_sims} shows the $\Delta \langle \log t \rangle_L$ variations against the $S/N$ of the original spectra. The lines represent the mean $\pm 1 \sigma$ range. Uncertainties do decrease with increasing $S/N$ in the SE0, SR1 and SR2 runs, but not in the C-runs. This is in fact not surpising, since the shape perturbations have the same amplitude for all spectra, independent of their $S/N$. The right-most panel illustrates the bias towards lower ages which results from the $A_V \ge 0$ barrier. 

We note in passing that, as hinted in Fig.\ \ref{fig:SN_sims}, the distribution of $S/N$ values in CALIFA 277 is skewed towards the target minimum $S/N$ of 20 stipulated in the spatial binning scheme explained in Paper I. This same threshold was applied to all other datacubes, so the uncertainty estimates presented here should be representative of the CALIFA data in general, even though they are based on a single galaxy.

\subsection{Exploring the statistics in IFS data: Uncertainties in radial profiles}
\label{sec:ErrorsRadialProfiles}

\begin{figure*}
\includegraphics[width=1\textwidth]{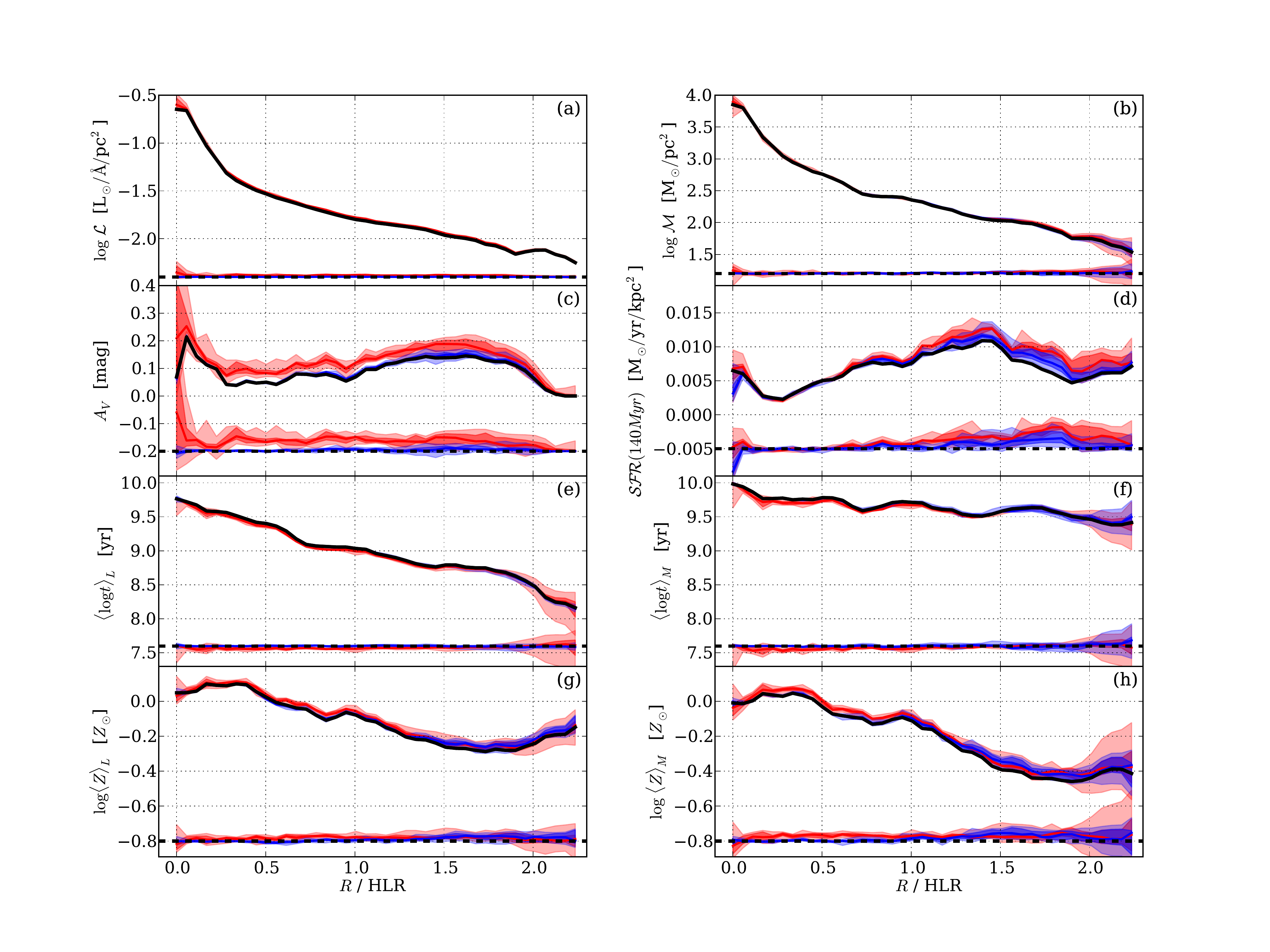}
\caption{Azymuthally averaged radial profiles of some global properties. 
(a) Dereddened luminosity surface density at 5635 \AA; (b) stellar mass surface density; (c) extinction; (d) SFR surface density over the last 140 Myr; (e) luminosity and (f) mass weighted mean log ages; (g) luminosity and (h) mass weighted mean metallicities. The original solution is shown as the solid black line, while those obtained from simulations are shown in colored shades. Light blue shades show the full range of values found among the 10 realizations of random noise (OR1 runs). The darker blue bands mark the range spanned by the 6 values closer to the median (marked with the blue solid line). OC002 simulations of color-change effects are plotted in red. The differences between the simulations and original profiles are shown at the bottom of each panel, with the zero level indicated by the dashed black line. The dispersion and offsets along these difference profiles give a measure of the uncertainties and biases involved in the radial profiles.}
\label{fig:ErrorsRadialProfiles} 
\end{figure*}

So far our error analysis has treated the datacube as a collection of unrelated spectra. Spectral synthesis does not care where a spectrum come from, so the positional information provided by IFS has no effect on the uncertainties for individual zones. However, averaging results obtained for different zones brings obvious statistical benefits, reducing uncertainties  in properties derived from individual spectra. 

Fig.\ 10 in Paper I shows azymuthally averaged profiles for a series of global properties. Despite the scatter at any given radius, gradients in age and metallicity were clearly detected in CALIFA 277. Fig.\ \ref{fig:ErrorsRadialProfiles} in this paper repeats that plot, but now  showing the radial profiles derived from the simulated spectra, so that we can evaluate the uncertainties in azymuthal averages. The original profiles are plotted in solid black, while colored shades represent the ranges spanned by OR1 (blue) and C002 (red) runs. Each panel also shows an (arbitrarely shifted) simulation minus original difference profile at its bottom. 

As expected, random noise does not change the radial profiles significantly. 
A small bias of about 0.02 mag is seen in $\overline{A_V}(R)$ for radii beyond $\sim 1$ Half Light Radius (HLR). This is again a side effect of the $A_V \ge 0$ constraint: When a spectrum originally fitted with $A_V \sim 0$ is perturbed, some will be refitted with $A_V > 0$, but none will be refitted with $A_V < 0$. The resulting bias propagates to minor offsets in other properties, like 
slightly younger ages (panels e and f). Overall, however, these are all tiny effects.

Color perturbations (red shaded regions) produce larger effects. The main effect is on the $\overline{A_V}(R)$ profile (panel c), where the typical offset matches the expected $\overline{\Delta}(A_V) = 0.05$ mag reported in Table \ref{tab:Stats_SimsK0277}. Leaving aside this bias, the $\overline{A_V}(R)$ variations are much smaller than those in individuals zones. For instance, considering all radial bins and all $10 \times 1638$ versions of the OC002 spectra, the rms in $\Delta(\overline{A_V})$ is only 0.05 mag, while individual zones are subjected to uncertainties of $\sigma(A_V) = 0.16$ mag (Fig.\ \ref{fig:DiffHists}). 

Larger than typical variations occur in the nucleus for the trivial reason that it comprises a single spaxel, so no actual statistics is done. More robust results for ``the nucleus'' require extending its definition to include more spaxels. For instance, P\'erez et al.\ (2013) define the nucleus as the inner 0.1 HLR, which in the case of CALIFA 277 comprises the central-most 9 spaxels. Uncertainties also increase in the outer regions ($R > 2$ HLR), induced by the smaller $S/N$ of the spectra and by the decrease in the number of zones compared to intermediate radii.

Fig.\ \ref{fig:ErrorsRadialProfiles} confirms the expected gain in dealing with averaged results. The simulations show that the uncertainties in the radial profiles are just too small to question the reality of the age and metallicity gradients in this galaxy. This conclusion holds even for our most pessimistic simulations (C004 and R3). This gain comes at the expense of a loss of spatial information---compressing the $x,y$ plane to a single dimension ($R$) in the example presented here. Other forms of spatial averaging can be envisaged, like isolating structural components (bulge, disk, bar, arms, inter-arm regions). From the results above, it is easy to see that, as long as enough zones are involved in the statistics, any averaging scheme should lead to small uncertainties.

\subsection{Uncertainties in Star Formation Histories}
\label{sec:ErrorsSFHs}

\starlight\ outputs SFHs as population vectors: $\vec{x}$ and $\vec{\mu}$. The first gives the fractional contribution of each base element to the total flux at the chosen reference wavelength (5635 \AA\ in our case), while the latter gives the corresponding mass fractions.  The 39 ages in our base\footnote{The base ages were chosen such that SSP spectra for ages in between $t_j$ and $t_{j+1}$ are adequately interpolated with the $j$ and $j+1$ SSPs (see Mateus et al.\ 2006 for further details).} sample the 1 Myr to 14 Gyr interval with a mean $\log t$ separation of just 0.1 dex. Individually, the strength of each component is not reliable (see Fig.\ 1 in Cid Fernandes et al.\ 2004 for an example), given the many ways that spectrally similar components can rearrange their $x_j$ values to produce a similar total spectrum. Yet, their combination in indices such as $\langle \log t \rangle_L$ produces robust quantities, as shown by the simulations.

None of the properties studied in the previous sections relies heavily on the details of the spectral decomposition performed by \starlight. Mean ages, for instance, are defined as the {\em first moment} of the age distributions encoded in  $\vec{x}$ and $\vec{\mu}$. By construction, however, SFHs do rely on higher order descriptions of the age distribution. This section aims to evaluate the uncertainties in our SFHs. Before doing so, we digress on the actual meaning of the 'H' in SFH. 

There is no standard semantics in what concerns the meaning of 'History' in spectral synthesis studies. On the contrary, the same term is used in very different ways in the literature, always qualitatively referring to some measure of the how star formation unfolded over time, but quantitatively varying from  estimates of  mean ages, to measures of the 'recent' star formation (where recent can mean anything between $\sim 10$ Myr and a few Gyr, depending on the tracers), and full $t$-by-$t$ descriptions with varying age resolution. Besides different time scales, different metrics can be used to quantify the SFH, like luminosity or mass associated to stars in a given age range, in differential or cumulative forms, etc. This diversity is particularly natural in the case of non-parametric codes such as \starlight. Because no pre-established functional forms for SFR$(t)$ are imposed, one has plenty of freedom to design different indices and functions to characterize the SFH. Paper I is itself an example of this diversity. We have presented there a series of SFH diagnostics obtained by manipulating the $\vec{x}$ and $\vec{\mu}$  vectors in different ways, always with the underlying philosophy of compressing the over-detailed output of \starlight\ into coarser but more robust descriptions of the SFH. 
 
In short, there are numerous alternative ways to describe a SFH. In what follows we evaluate the uncertainties in a Paper I-inspired selection of SFH-descriptions. Following the sequence of our analysis of global properties, we first present results for uncertainties in individual zones (\S\ref{sec:ErrorsSFH_IndividualZones}) and then for spatially averaged SFHs (\S\ref{sec:ErrorsSFH_SpatialAverages}).

\subsubsection{SFHs for individual zones}
\label{sec:ErrorsSFH_IndividualZones}

\begin{table*}
\begin{centering}
\begin{tabular}{lrrrrrrr}
\multicolumn{8}{c}{Uncertainties in light fraction in age groups: $\overline{\Delta x} \pm \sigma_{\Delta x}$ [\%]} \\ \hline
%
%
%
%
$\log t$ range    &    OE0    &    OR1    &    OR2    &    OR3    &    OC001    &    OC002    &    OC004   \\
\hline
6.0$\rightarrow$8.2  &   0.0   $\pm$ {\bf    1.9  }    &   0.1   $\pm$ {\bf    3.3  }    &   -0.0   $\pm$ {\bf    4.9  }    &   -0.0   $\pm$ {\bf    6.0  }    &   0.1   $\pm$ {\bf    2.2  }    &   0.6   $\pm$ {\bf    2.7  }    &   1.6   $\pm$ {\bf    4.2  }  \\
8.2$\rightarrow$9.2  &   0.0   $\pm$ {\bf    4.5  }    &   -0.1   $\pm$ {\bf    9.3  }    &   0.7   $\pm$ {\bf    14.9  }    &   1.2   $\pm$ {\bf    18.9  }    &   0.7   $\pm$ {\bf    6.8  }    &   4.3   $\pm$ {\bf    14.7  }    &   9.4   $\pm$ {\bf    23.2  }  \\
9.2$\rightarrow$10.2  &   -0.1   $\pm$ {\bf    4.0  }    &   -0.2   $\pm$ {\bf    8.7  }    &   -1.0   $\pm$ {\bf    14.0  }    &   -1.7   $\pm$ {\bf    17.9  }    &   -0.8   $\pm$ {\bf    6.4  }    &   -5.0   $\pm$ {\bf    14.8  }    &   -11.6   $\pm$ {\bf    23.9  }  \\
\hline
%
%
%
6.0$\rightarrow$7.5   &   -0.0   $\pm$ {\bf    1.4  }    &   -0.2   $\pm$ {\bf    3.0  }    &   -0.4   $\pm$ {\bf    4.7  }    &   -0.6   $\pm$ {\bf    5.8  }    &   0.0   $\pm$ {\bf    1.6  }    &   0.3   $\pm$ {\bf    2.4  }    &   0.8   $\pm$ {\bf    4.0  }  \\
7.5$\rightarrow$8.5   &   -0.4   $\pm$ {\bf    3.1  }    &   -0.7   $\pm$ {\bf    5.9  }    &   -1.3   $\pm$ {\bf    8.4  }    &   -1.7   $\pm$ {\bf    10.1  }    &   -0.4   $\pm$ {\bf    3.5  }    &   -0.4   $\pm$ {\bf    4.8  }    &   -0.1   $\pm$ {\bf    7.2  }  \\
8.5$\rightarrow$9.3   &   0.0   $\pm$ {\bf    4.4  }    &   0.0   $\pm$ {\bf    9.4  }    &   0.6   $\pm$ {\bf    15.1  }    &   1.2   $\pm$ {\bf    19.2  }    &   0.7   $\pm$ {\bf    6.7  }    &   4.5   $\pm$ {\bf    14.7  }    &   9.8   $\pm$ {\bf    23.3  }  \\
9.3$\rightarrow$9.7   &   -0.1   $\pm$ {\bf    5.5  }    &   -0.7   $\pm$ {\bf    12.8  }    &   -2.3   $\pm$ {\bf    19.6  }    &   -4.0   $\pm$ {\bf    23.8  }    &   -0.6   $\pm$ {\bf    7.7  }    &   -3.2   $\pm$ {\bf    13.2  }    &   -7.4   $\pm$ {\bf    18.2  }  \\
9.7$\rightarrow$10.2   &   -0.1   $\pm$ {\bf    3.4  }    &   0.1   $\pm$ {\bf    10.4  }    &   0.4   $\pm$ {\bf    16.4  }    &   1.0   $\pm$ {\bf    20.1  }    &   -0.4   $\pm$ {\bf    5.6  }    &   -2.3   $\pm$ {\bf    11.0  }    &   -5.0   $\pm$ {\bf    15.4  }  \\
\hline
\end{tabular}
\end{centering}
\caption{As Table \ref{tab:Stats_SimsK0277} but with the statistics of the light fractions in age groups for our different simulations. simulations. For each age group the table lists the mean simulated minus original difference ($\overline{\Delta}$) and its standard deviation ($\sigma_\Delta$, in bold), both in percentages. 
The top part of the table corresponds to the Young, Intermediate and Old groups description discussed in the text and Fig.\ \ref{fig:dx_3groups}, while the bottom one is for 5-groups used in Fig.\ \ref{fig:dx_5groups}. The R1 and C002 columns are those applicable to CALIFA data.}
\label{tab:Stats_XAgeGroups}
\end{table*}

\begin{table*}
\begin{centering}
\begin{tabular}{lrrrrrrr}
\multicolumn{8}{c}{Uncertainties in $\log$ mass in age groups: $\overline{\Delta \log M^\prime} \pm \sigma_{\Delta \log M^\prime}$ [dex]} \\ \hline
%
%
%
$\log t$ range    &    OE0    &    OR1    &    OR2    &    OR3    &    OC001    &    OC002    &    OC004   \\
\hline
6.0$\rightarrow$8.2    &   0.00   $\pm$ {\bf    0.14  }    &   -0.00   $\pm$ {\bf    0.21  }    &   -0.02   $\pm$ {\bf    0.30  }    &   -0.03   $\pm$ {\bf    0.37  }    &   0.00   $\pm$ {\bf    0.16  }    &   0.02   $\pm$ {\bf    0.20  }    &   0.06   $\pm$ {\bf    0.29  }  \\
8.2$\rightarrow$9.2   &   0.00   $\pm$ {\bf    0.12  }    &   -0.01   $\pm$ {\bf    0.25  }    &   -0.00   $\pm$ {\bf    0.37  }    &   -0.00   $\pm$ {\bf    0.44  }    &   0.01   $\pm$ {\bf    0.18  }    &   0.08   $\pm$ {\bf    0.28  }    &   0.15   $\pm$ {\bf    0.36  }  \\
9.2$\rightarrow$10.2  &   -0.00   $\pm$ {\bf    0.05  }    &   -0.00   $\pm$ {\bf    0.12  }    &   -0.01   $\pm$ {\bf    0.19  }    &   -0.02   $\pm$ {\bf    0.24  }    &   -0.01   $\pm$ {\bf    0.08  }    &   -0.04   $\pm$ {\bf    0.20  }    &   -0.05   $\pm$ {\bf    0.33  }  \\
\hline
%
%
%
6.0$\rightarrow$7.5   &   -0.00   $\pm$ {\bf    0.11  }    &   -0.02   $\pm$ {\bf    0.18  }    &   -0.05   $\pm$ {\bf    0.27  }    &   -0.07   $\pm$ {\bf    0.32  }    &   -0.00   $\pm$ {\bf    0.12  }    &   0.01   $\pm$ {\bf    0.14  }    &   0.02   $\pm$ {\bf    0.18  }  \\
7.5$\rightarrow$8.5   &   -0.00   $\pm$ {\bf    0.22  }    &   0.01   $\pm$ {\bf    0.35  }    &   0.06   $\pm$ {\bf    0.43  }    &   0.09   $\pm$ {\bf    0.49  }    &   0.00   $\pm$ {\bf    0.24  }    &   0.03   $\pm$ {\bf    0.29  }    &   0.12   $\pm$ {\bf    0.39  }  \\
8.5$\rightarrow$9.3   &   0.00   $\pm$ {\bf    0.12  }    &   -0.00   $\pm$ {\bf    0.24  }    &   0.01   $\pm$ {\bf    0.35  }    &   0.03   $\pm$ {\bf    0.41  }    &   0.01   $\pm$ {\bf    0.18  }    &   0.08   $\pm$ {\bf    0.28  }    &   0.16   $\pm$ {\bf    0.36  }  \\
9.3$\rightarrow$9.7   &   0.00   $\pm$ {\bf    0.14  }    &   -0.01   $\pm$ {\bf    0.25  }    &   -0.02   $\pm$ {\bf    0.34  }    &   -0.03   $\pm$ {\bf    0.39  }    &   -0.01   $\pm$ {\bf    0.17  }    &   -0.03   $\pm$ {\bf    0.24  }    &   -0.04   $\pm$ {\bf    0.31  }  \\
9.7$\rightarrow$10.2   &   -0.00   $\pm$ {\bf    0.11  }    &   -0.01   $\pm$ {\bf    0.24  }    &   -0.00   $\pm$ {\bf    0.34  }    &   0.02   $\pm$ {\bf    0.38  }    &   -0.01   $\pm$ {\bf    0.15  }    &   -0.03   $\pm$ {\bf    0.27  }    &   -0.00   $\pm$ {\bf    0.36  }  \\
\hline
\end{tabular}
\end{centering}
\caption{As Table \ref{tab:Stats_XAgeGroups} but for the (log) initial masses associated to stellar populations in different age groups.}
\label{tab:Stats_logMAgeGroups}
\end{table*}

\begin{figure}
\includegraphics[width=0.5\textwidth]{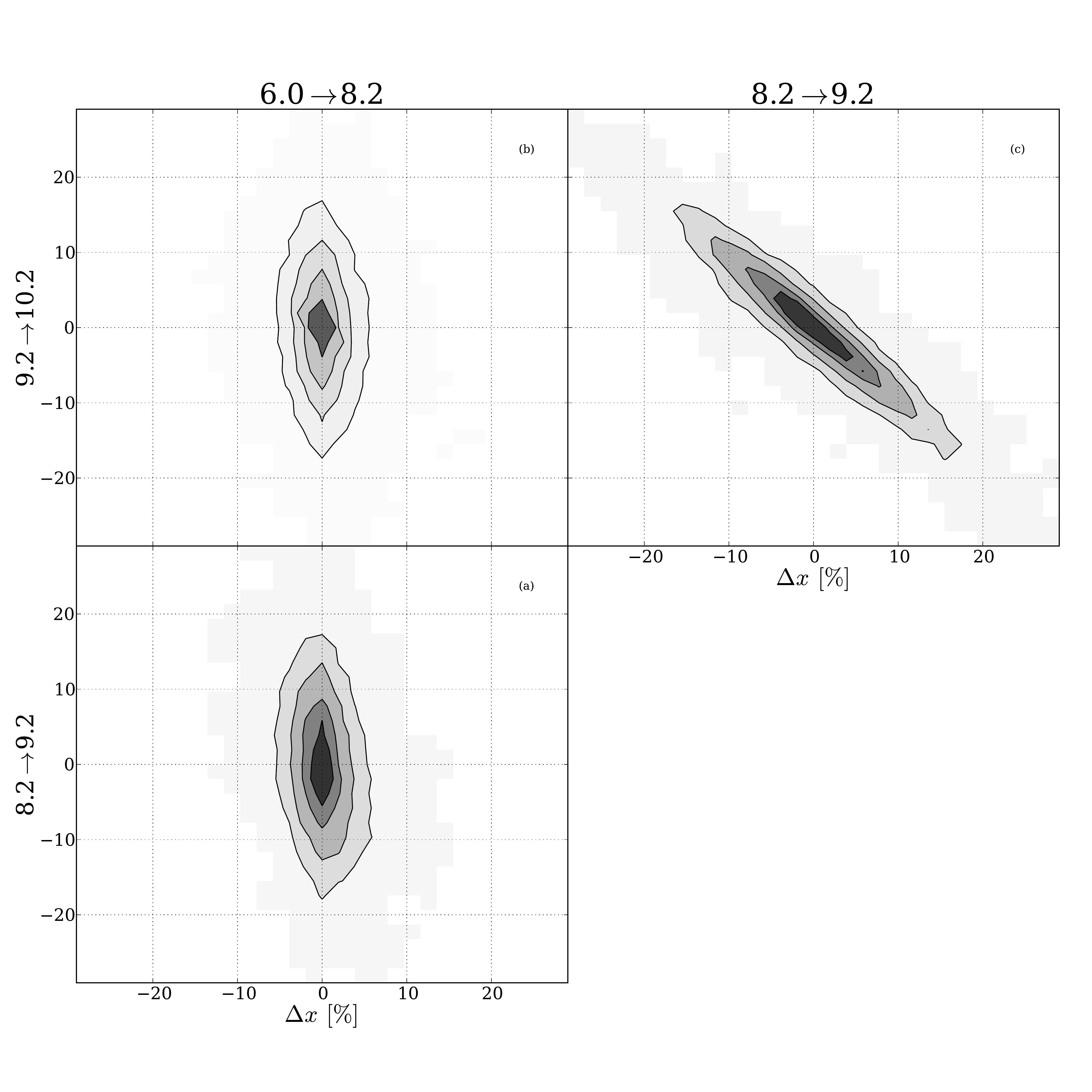}
\caption{Variations in the light fractions in Young, Intermediate and Old age groups (spanning $\log t =$ 6--8.2, 8.2--9.2 and 9.2--10.2, respectively) for OR1 simulations. Contours are drawn at 20, 40, 60 and 80\% of enclosed points.
}
\label{fig:dx_3groups} 
\end{figure}

\begin{figure}
\includegraphics[width=0.5\textwidth]{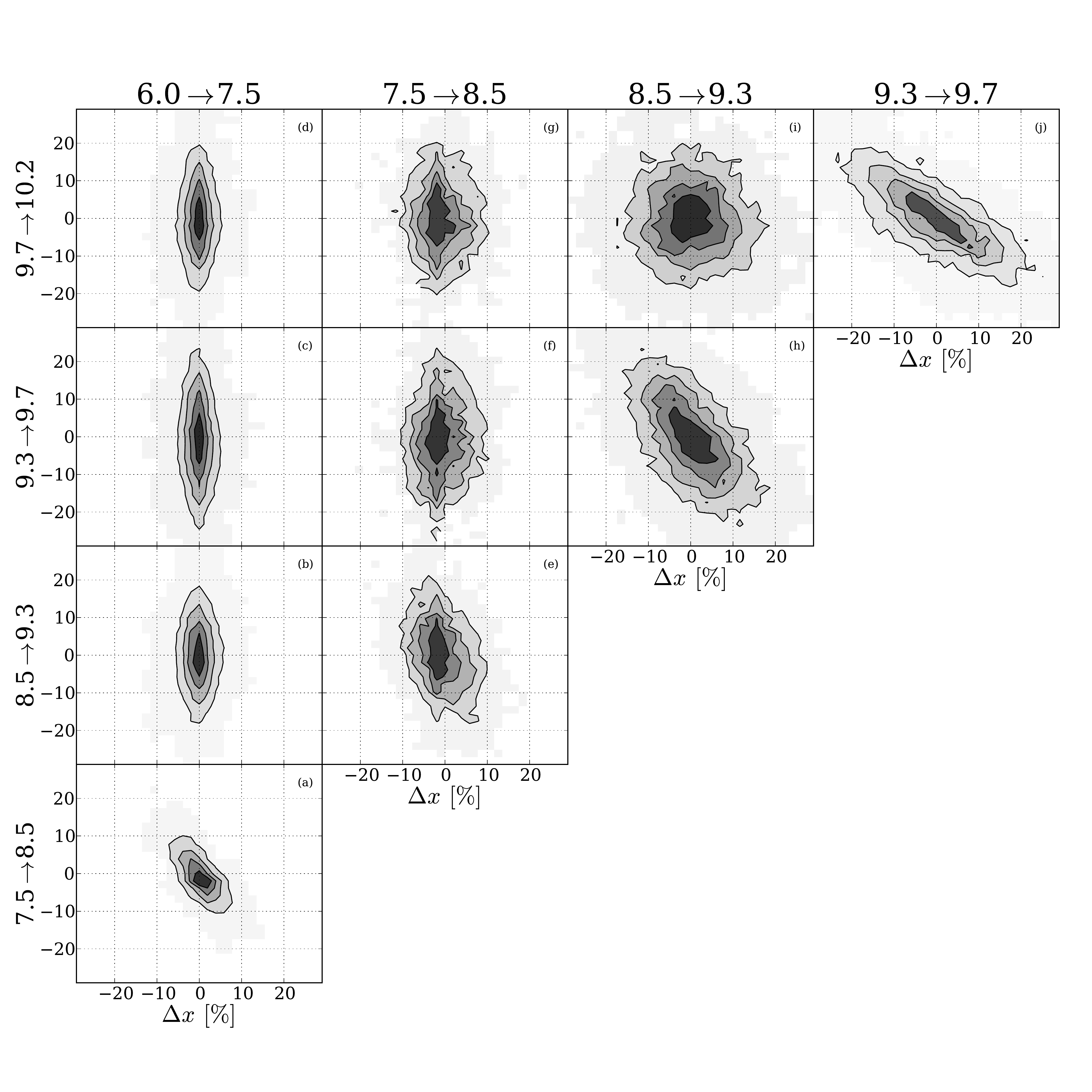}
\caption{As Fig.\ref{fig:dx_3groups}, but for a finer graded description of the SFH in terms of 5 age groups.
}
\label{fig:dx_5groups} 
\end{figure}

\begin{figure}
\includegraphics[width=0.5\textwidth]{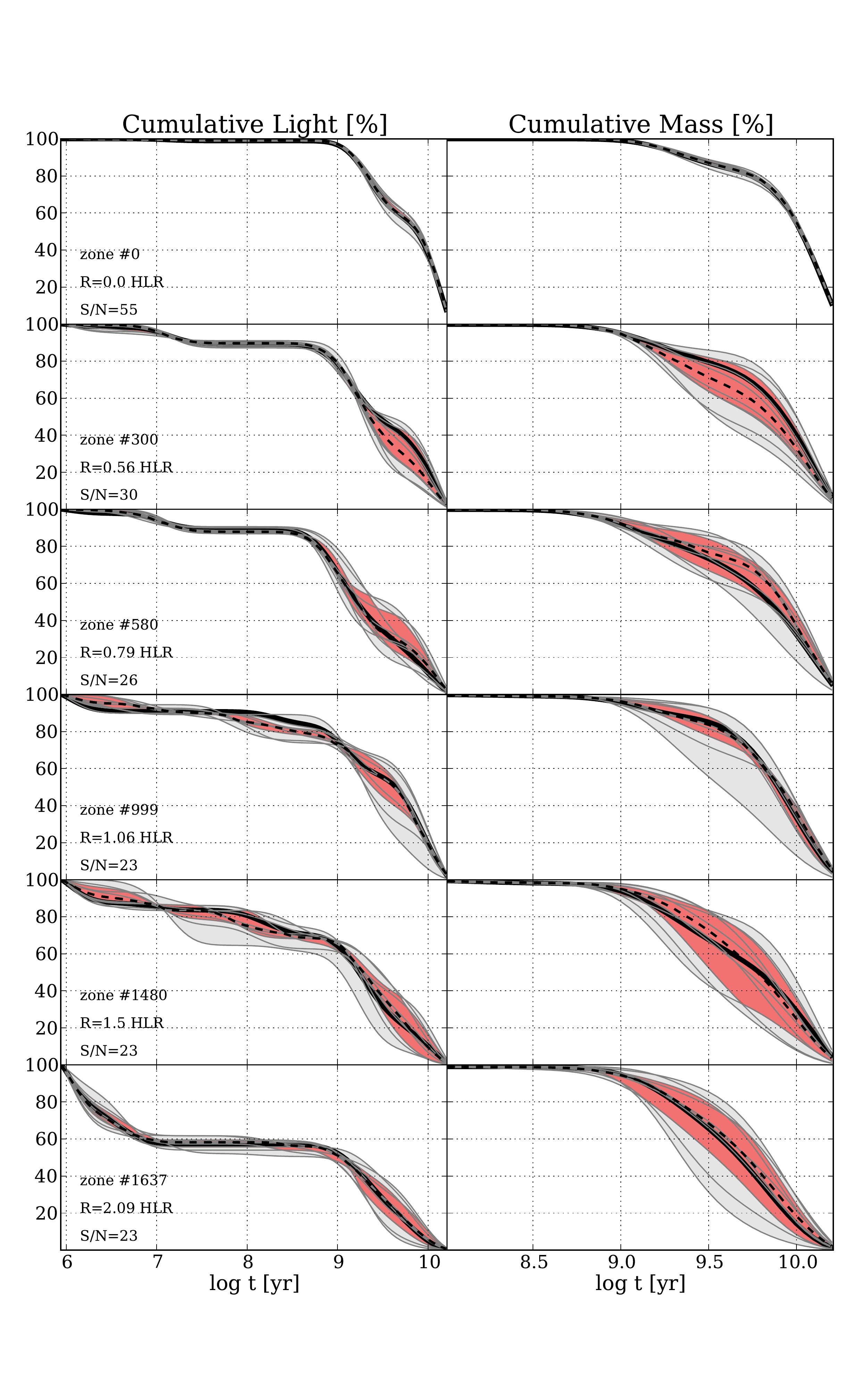}
\caption{SFHs of six different spatial zones, as represented by their cumulative light (left) and mass (right) distributions. Each panel shows 11 SFHs. The original one is plotted in thick black lines. Grey lines show the SFHs for the 10 realizations of OR1 simulations, with a grey shaded band tracing the full range of values. The median SFH is drawn as a dashed line, while the red band marks the range spanned by the 6 runs closer to the median. The zone number, distance to the nucleus and $S/N$ are labeled in the left panels.
}
\label{fig:ExCumulSFHs} 
\end{figure}

One way to characterize SFHs is to compress the population vectors into  age ranges, an approach which dates back to the early days of spectral synthesis (Bica 1988). In the context of \starlight, Cid Fernandes et al.\ (2004 and 2005) showed that a robust summary of the SFH can be obtained by binning $\vec{x}$ into young (Y), intermediate (I) and old (O)  groups. This $\vec{x} \rightarrow (x_Y,x_I,x_O)$ compression scheme was employed to produce the maps shown in Figs.\ 8 and  9 of Paper I, where these age groups were separated by frontiers placed at $t = 0.14$ and 1.4 Gyr. 

We have used our simulations to assess the uncertainties in these fractions. The top three rows in Table \ref{tab:Stats_XAgeGroups} list the O-runs statistics of $\Delta x_Y$ , $\Delta x_I$ and $\Delta x_O$, where, as before, $\Delta \equiv$  Simulated $-$ Original value. The table shows that random noise at the one-$\sigma$ level (R1 runs) induces uncertainties in $x_Y$, $x_I$ and $x_O$ of $\sim 3$, 9 and 9\% respectively\footnote{Notice that \% here does {\em not} mean a relative deviation, but the very units of the $x_Y$, $x_I$ and $x_O$ fractions.}. Larger uncertainties ($\sim 15$\%) in $x_I$ and $x_O$ apply for shape-changing perturbations at the level expected for CALIFA (OC002 runs). Part of this increase is due to aformentioned $A_V \ge 0$ effect, which forces low $A_V$ zones which are perturbed to bluer spectra to become younger to compensate the inability to reduce $A_V$ below 0. This is also reflected in the biases, with $\overline{\Delta}(x_I) \sim - \overline{\Delta}(x_O) \sim +5$\%. At any rate, comparing these uncertainties with  $(\overline{x_Y} , \overline{x_I} , \overline{x_O}) = (15 , 25 , 60)$\%, the average values over all zones in CALIFA 277, one concludes that this compact description of the SFH is reasonably robust.

Fig.\ \ref{fig:dx_3groups} show how the variations in these light fractions correlate with each other. $\Delta x_O$ anti-correlates very strongly with $\Delta x_I$, which is not suprising given the $x_Y + x_I + x_O = 100\%$ constraint and the fact that these are the age groups which carry most of the light in CALIFA 277. The Y group, on the other hand, is roughly independent of the others. 

Fig.\ \ref{fig:dx_5groups} shows OR1-based results for a finer-graded description using 5 age groups, defined by $\log t \,{\rm [yr]} = 6$--7.5, 7.5--8.5, 8.5--9.3, 9.3--9.8 and 9.8--10.2. The anti-correlations among adjacent age groups are visible along the diagonal. The $\sigma_{\Delta x}$ values are 3, 6, 9, 13 and 10\%
(Table \ref{tab:Stats_XAgeGroups}), for mean fractions of $\overline{x_t} =13$, 3, 24, 34 and 26\%, respectively. This more detailed description shortens the difference between $\sigma_{\Delta x}$ values and the corresponding $\overline{x}$. The uncertainties are therefore larger, as expected, but these higher resolution SFHs are still useful, specially if used in a statistical way.

As the name implies, \starlight\ fits light, so $\vec{x}$-based SFH descriptions are the natural way to analyse its results. Yet, it is the stellar mass which carries the astrophysical information on the galaxy history. One can obviously group $\vec{\mu}$ elements just as we did for $\vec{x}$, but while light fractions in different age groups can be directly intercompared, mass fractions cannot. With a factor of $\sim 1000$ decrease in light-to-mass ratio from the youngest to the oldest elements in the base, wildly different mass fractions, say, $\mu_Y \sim 1$\% and $\mu_O \sim 99\%$, can be equally relevant for the fits.

To evaluate uncertainties in the masses associated to different age groups it is better to use the masses themselves instead of mass-fractions. Table \ref{tab:Stats_logMAgeGroups} reports the uncertainties in $\log M^\prime$, where the prime denotes the initial stellar mass, uncorrected for the mass returned to the interstellar medium by stellar evolution. The table reports the results of the simulations for both the (Y,I,O) and the 5-age groups descriptions of the SFH discussed above. Again, the fact that we are now exploring higher moments of the age distribution implies larger uncertainties than for global properties. A column-by-column comparison with the $\Delta \log M^\prime$ statistics in Table \ref{tab:Stats_SimsK0277} confirms this. For instance, while the total masses have uncertainties of just 0.08 dex in OR1 runs, the masses in age-groups are typically 3 times more uncertain, an unavoidable but still reasonable price to pay for a time-resolved description of results of the spectral synthesis.

Further subdivisions lead to a quasi continuous regime which is better handled in ways other than age-binning, like cumulating $\vec{x}$ or $\vec{\mu}$. Fig.\ \ref{fig:ExCumulSFHs} illustrates this. The left panels show the cumulative light fraction, shown after a cosmetic smoothing by a FWHM $= 0.2$ dex gaussian in $\log t$. Six individual zones, including the nucleus (top) and the last one (bottom), were chosen to exemplify this alternative description of SFHs and its uncertainties. The thick solid line traces the original \starlight\ solution, while thin grey lines trace the solutions obtained for each of the 10 OR1 perturbations, whose median is drawn as a dashed line. The red-shaded bands mark the range spanned by the 6 SFHs closer to the median at each $t$. The right panels show the corresponding mass assembly histories, all scaled to 100\% at $t = 0$. The six zones in Fig.\ \ref{fig:ExCumulSFHs} cover different distances from the nucleus, $S/N$ and  SFHs, thus providing representative examples of \starlight\ results for CALIFA data. 

Except for the nucleus, where the perturbed spectra and the associated SFHs are nearly indistinguishable from the original ones because of the very high $S/N$, there is a clear spread in cumulative light and mass distributions. The dispersion is larger for ages larger than a few Gyrs, which contain a substantial fraction of the light and most of the mass. The mild spectral evolution in this age range (Conroy 2013) explains the larger uncertainties, as components can be interchanged with relatively little variation in the total spectrum. Gauging the uncertainties in the cumulative SFHs by the width of the red band, we find it to be of the order of 10--20\%, comparable to those reported in Tables \ref{tab:Stats_XAgeGroups} and \ref{tab:Stats_logMAgeGroups} for age grouped quantities.

Despite uncertainties, the example SFHs in Fig.\ \ref{fig:ExCumulSFHs} reveal a gradual shift towards younger ages as one moves to larger radii. This trend is the higher order description of clear mean age radial gradients previously shown in  panels e and f of Fig.\ \ref{fig:ErrorsRadialProfiles}, where results for individual zones at same $R$ were averaged.

\subsubsection{Spatially averaged SFHs}
\label{sec:ErrorsSFH_SpatialAverages}

\begin{figure}
\includegraphics[width=0.5\textwidth]{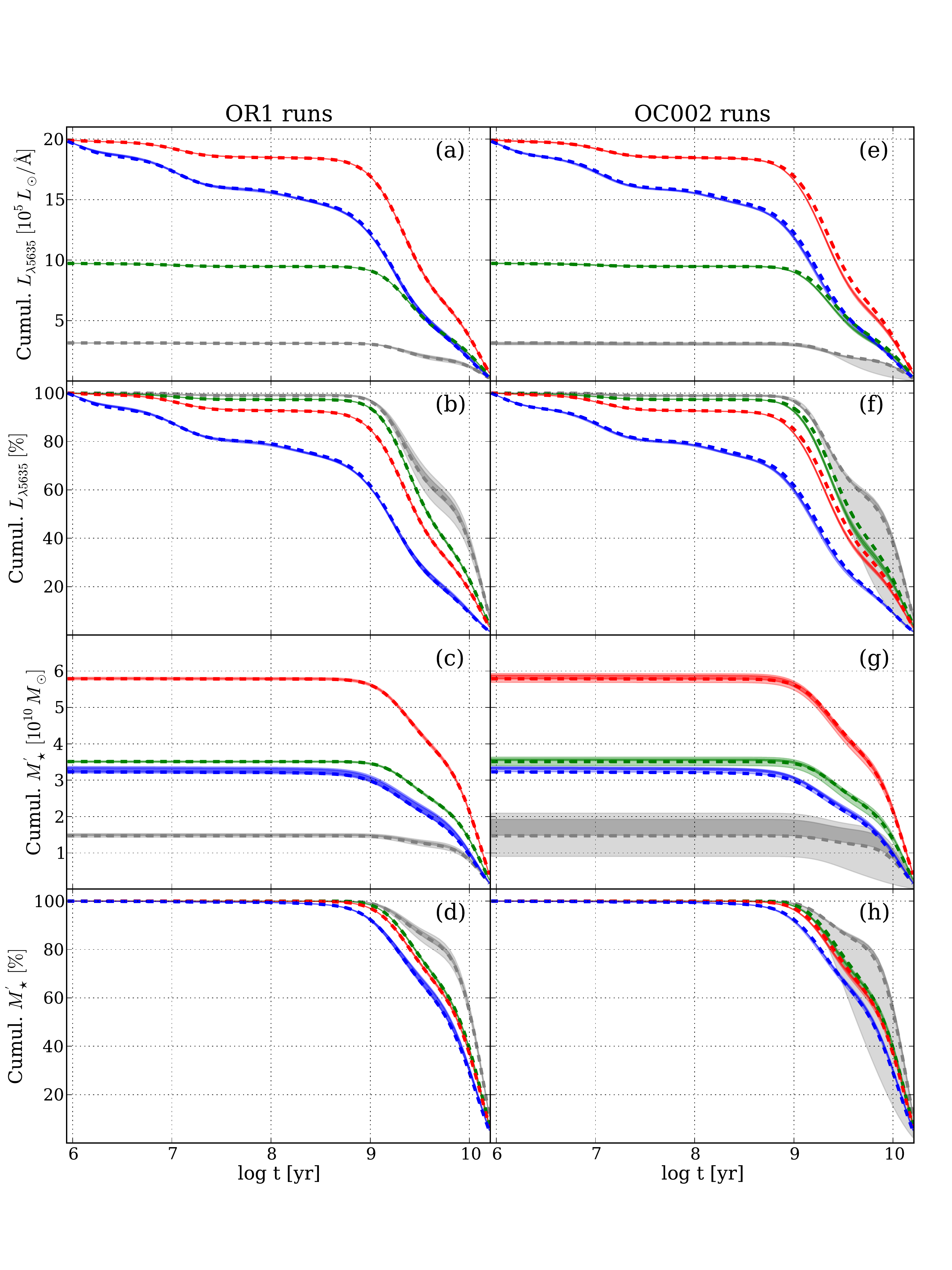}
\caption{Spatially averaged cumulative light and mass histories and their uncertainties for four regions: Red, blue, green and grey colors correspond to regions $R > 1$, $R < 1$, $R < 0.5$ HLR and the nucleus, respectively. Left and right panels show OR1 and OC002 simulations, respectively. The top four panels show cumulative light distributions, in both absolute (a, e) and relative (b, f) units. Similarly, the bottom four panels show the correponding mass distributions. In each panel and for each radial region the original solution is drawn as a dashed line, while results of the simulations are shown with shaded bands, hardly visible except for the nuclear curves (grey). Light shaded bands show the ranges spanned by all 10 perturbations, and darker shades indicate the range spanned by the 6 versions closer to the median. The nuclear curves in panels a, c, e and g contain much less light and mass than the others, and so were multiplied by 20 for clarity.
}
\label{fig:Errors_RadRegions_SFHs} 
\end{figure}

In analogy with \S\ref{sec:ErrorsRadialProfiles}, uncertainties in the SFHs of individual zone spectra should be greatly reduced when averaging over spatial regions comprising many zones. This is confirmed in Fig.\ \ref{fig:Errors_RadRegions_SFHs}, which show cumulative light and mass distributions (in both absolute and relative units) for four spatial regions: The nucleus (grey), $R < 0.5$ (green), $R < 1$ (red) and $R > 1$ HLR (blue). The smoothing here is by FWHM $= 0.5$ dex in $\log t$ for compatibility with Fig.\ 11 of Paper I. OR1 runs are shown in the left panels, and OC002 on the right.

The statistical damping of the uncertainties is so strong that one can hardly notice that, as in Fig.\ \ref{fig:ExCumulSFHs}, shaded bands are used to show the vertical ranges spanned by the simulations. The bands are so narrow that they look like thick lines. The ``exception'' is the nucleus (grey), where the uncertainties are visible for the simple reason that, unlike for other regions, it does not involve any actual spatial averaging because it comprises a single spaxel. 

Close inspection of the right panels reveals that the shape-changing simulations tend to have SFHs slightly skewed towards younger ages than SFHs inferred from spectral fits of the original data (marked by dashed lines). This is again the effect of some perturbations making low $A_V$ zones too blue to be fitted with $A_V \ge 0$, so \starlight\ compensates it with lower ages. 

In conclusion, spatial averaging effectively eliminates the dispersions seen in the SFHs of individual zones. This remains valid even for simulations where the level of perturbations are  higher than those expected for CALIFA spectra.

\subsection{Summary of the simulations}
\label{sec:ErrorsConclusion}

The overall message from these experiments is that degeneracies inherent to stellar populations and uncertainties in the data have a limited impact upon our analysis.  For the level of uncertainties expected in CALIFA, here represented by the R1 and C002 runs, the simulations yield acceptably small uncertainties in global properties. More precisely, mean ages and metallicities have one sigma variations of $\sim 0.1$ dex when weighted in luminosity and $\sim 0.15$ weighting by mass, while stellar masses are good to $\sim 0.1$ dex (round numbers), with negligible biases. Uncertainties in $A_V$ are $\sim 0.07$ mag if only random noise is considered, increasing to 0.16 mag when shape-related uncertainties are accounted for. In the latter case, a bias of $\overline{\Delta} A_V \sim 0.06$ mag appears as a side-effect of forbidding negative $A_V$. Being a higher order product of the spectral synthesis, SFHs are naturally less strongly constrained than global properties, but the age-grouped and cumulative descriptions explored above were found to be relatively stable against random and spectral shape perturbations. 

It is a consensus in the field of spectral synthesis that the results are best used in a statistical sense (e.g., Panter et al.\ 2006), and the present study endorses this view. First, averaging reduces uncertainties, as confirmed by our experiments with radial profiles (Fig.\ \ref{fig:ErrorsRadialProfiles}) and spatially averaged SFHs (Fig.\ \ref{fig:Errors_RadRegions_SFHs}), which showed that the results are stable against the kinds of uncertainties studied here. Second, and more importantly, large samples allow for {\em comparative} studies, which can reveal astrophysically valuable trends while mitigating worries about data or method related systematics affecting the absolute values. This point has been abundantly explored in spectral synthesis studies of SDSS galaxies, where inequivocal relations between, say, age, mass and metallicity, have been identified in spite of the uncertainties in the properties derived for individual galaxies. The analogy with IFS data is clear. In IFS each galaxy is a large sample by itself, and the very goal of IFS is to compare properties in one place against those somewhere else. 

This concludes our study of data and method related uncertainties. This whole study was based on a single set of SSP models, a fundamental prior in the analysis. We now turn to the uncertainties resulting from this choice.

\section{Uncertainties associated to evolutionary synthesis models: An empirical evaluation}
\label{sec:SSPtests}

\begin{table*}
\centering
\caption{Sets of SSP models used for spectral fitting}
\begin{tabular}{@{}lccccccrc@{}} \hline
Model-set &     Code  &  Isochrones  & IMF  &  Library  &  Metallicities  &  $N_Z \times N_t$  &  Ages [Gyr]\\  \hline
{\it GM}     & SED@ + Vazdekis    & Padova2000 & Salpeter  & {\sc granada}  + MILES & 0.004, 0.008, 0.019, 0.033  &   4$\times$39 & 0.001--14\\ 
{\it CB}    & Galaxev  & Padova1994 & Chabrier  & {\sc granada}  + MILES   &  
0.004, 0.008, 0.02, 0.05 &  4$\times$40 & 0.001--14\\
{\it BC}   &  Galaxev  & Padova1994 & Chabrier  & STELIB  &  0.004, 0.008, 0.02, 0.05 &  4$\times$40 & 0.001--14\\ 
\hline
\label{tab:STPop_Models_Summary}
\end{tabular}
\end{table*}

SSP spectra are the key astrophysical ingredient in any spectral synthesis analysis, the dictionary which translates stellar properties to spectroscopic information,  which can then be compared to data through different methods to extract physical properties and SFHs. The computation of model SSP spectra combines an assumed initial mass function (IMF) with stellar evolution theory (isochrones) and a library to provide the spectra of stars with  temperature, gravity and chemical composition matching the values dictated by the evolutionary tracks. In practice, this algebric exercise stumbles upon a series of issues which directly affect the predicted SSP spectra.

First, despite its (well deserved) status as a founding stone of modern astrophysics, stellar evolution is still an open business. Rotation in massive stars (Levesque et al.\ 2012), rapid phases during the post main sequence evolution of intermediate mass stars (Maraston 2005), and binarity (Li et al.\ 2013)  are examples of phenomena which affect the radiative output of stellar populations, but are still at the forefront of research, and thus prone to uncertainties. A practical example of how incomplete treatment of evolutionary phases affects spectral synthesis analysis is discussed by Koleva et al.\ (2008), Cid Fernandes \& Gonz\'alez Delgado (2010) and Ocvirk (2010). Using different codes, these studies find that spectral synthesis of old Milky Way and LMC globular clusters suggests the presence of very young stellar populations accounting for some 10--20\% of the optical light, an astrophysically absurd result which also happens in some old, ``red and dead'' galaxies (Cid Fernandes et al.\ 2011). These ``fake young bursts'' (Ocvirk 2010) are in fact byproducts of inadequate modeling of the horizontal branch (and/or blue straglers, Cenarro et al.\ 2008). Seeking for a best match, the spectral synthesis uses young hot stars to replace the old hot ones missing in the models.

Secondly, spectral libraries, be they empirical or theoretical, have their own limitations (Martins \& Coelho 2007). The widely used Bruzual \& Charlot (2003; BC03) models, for instance, are based on the STELIB library (Le Borgne et al.\ 2003), a major advance in its days, but limited to 249 stars, some of which with severe spectral gaps. Its relatively poor coverage of the stellar parameter space propagates to the predicted SSP spectra, with collateral effects on the results of a spectral synthesis analysis. Koleva et al.\ (2008), for instance, noticed that, because of the lack of truly very metal rich stars in the library, BC03 spectra for $2.5 Z_\odot$ SSPs behave like $\sim Z_\odot$ models of larger age. In short, this too is an evolving field, undergoing constant revisions.

In this unstable scenario, it is relevant to evaluate to which extent the results of our spectral synthesis analysis depend on the choice of SSP models. To this end, we have performed spectral fits using three bases extracted from different sets of evolutionary synthesis models, listed in Table \ref{tab:STPop_Models_Summary}. We stress that this is by no means an exaustive exploration of this issue, but it suits our goal of estimating uncertainties related to this choice. \S\ref{sec:Bases} describes the sets of models to be compared. The CALIFA data used in this study are described in \S\ref{sec:Data4SSPsTest}. Comparisons of the properties obtained with different sets of SSPs are presented in \S\ref{sec:BaseComparison}. Fit quality indicators and spectral residuals are discussed in \S\ref{sec:SpectralResiduals}, while \S\ref{sec:Comparison2SDSS} compares residuals in CALIFA and SDSS spectra.

\subsection{SSP spectral models}
\label{sec:Bases}

Base {\it GM} is a combination of publicly available SSP spectra from Vazdekis et al.\ (2010)\footnote{miles.iac.es}, which start at an age of 63 Myr, with the Gonz\' alez Delgado et al.\ (2005)\footnote{www.iaa.csic.es/$\sim$mcs/sed@} models for younger ages. The former are based on stars from the MILES library (S\' anchez-Bl\' azquez et al. 2006), while the latter relies on the synthetic stellar spectra from the {\sc granada}  library (Martins et al.\ 2005). Only minor adjustments were needed to match these two sets of models. First, we smoothed the {\sc granada} models to the spectral resolution of MILES (2.5 \AA\ FWHM, cf.\ Beifiori et al.\ 2011 and Falc\'on-Barroso et al.\ 2011), and then multiplied them by a factor of 1.05, estimated from a direct comparison of the predicted spectra in the 63 Myr to 1 Gyr range, where these models agree very well in both continuum shape and absorption features. These independently derived SSP spectra complement each other, providing a base suitable for spectral fitting analysis of galaxies of all types.

The evolutionary tracks in base {\it GM} are those of Girardi et al. (2000), except for the youngest ages (1 and 3 Myr), which are based on Geneva tracks (Schaller et al.\ 1992; Schaerer et al.\ 1993a,b, Charbonnel et al.\ 1993). The specific subset of SSPs used in the spectral fits comprises 39 ages between 1 Myr and 14 Gyr. Four metallicities are included: $Z = 0.2$, 0.4, 1 and 1.5 solar. The Salpeter IMF is adopted. Like the other bases discussed next, {\em GM} does not cover adequately the blue horizontal branch, both because it does not reach very low $Z$ and because, as thoroughly discussed by Ocvirk (2010), these models perform poorly in this phase anyway.

Base {\it CB} is an updated version of the BC03 models (Charlot \& Bruzual 2007, private communication), replacing STELIB by spectra from the MILES and {\sc granada} libraries. The IMF is that of Chabrier (2003), and the evolutionary tracks are those by Alongi et al.\ (1993), Bressan et al.\ (1993), Fagotto et al.\ (1994a, 1994b), and Girardi et al.\ (1996), collectively labeled 'Padova 1994' by BC03. $N_\star = 160$ SSPs comprising 40 ages in the 1 Myr to 14 Gyr range and 4 metallicities (0.2, 0.4, 1 and 2.5 $Z_\odot$) were chosen for the spectral fits.

Finally, base {\it BC} is built from the ``standard'' BC03 models, whose ingredients are identical to those of base {\it CB} except for the stellar library, which is STELIB.  Base {\it BC} allows us to check if differences between results obtained with bases {\it GM} and {\it CB} are associated to  isochrones and/or IMF instead of stellar libraries. 

These three bases cover the same age range, and all have the same number of metallicities. All bases start at $Z = 0.004$ ($0.2 Z_\odot$)\footnote{We exclude lower metallicity models to avoid SSPs that are not so well covered by the stellar libraries. Base {\it GM} base could include SSPs at $Z = 0.001$, but we do not do so because this metallicity is not available in the {\it CB} and {\it BC} models.}, but {\it GM} stops at $1.5 Z_\odot$, while both {\it CB} and {\it BC} go up to $2.5 Z_\odot$. Because they are both based on stars not far from the Sun, neither MILES nor STELIB actually contain many stars as metal rich as $2.5 Z_\odot$, so the latter models should be interpreted with care. On the other hand, the $Z \le 1.5 Z_\odot$ limit of base {\it GM} may be too low to accomodate the metal rich inner regions of galaxies, possibly leading to saturation effects. Base {\it GM} also differs from the others in IMF and isochrones. The differences in opacities and equation of state between Padova 2000 ({\it GM}) and 1994 ({\it CB} and {\it BC}) tracks produce somewhat warmer (hence bluer) stars in the red giant branch in the former than in the latter. In principle, one may thus expect to obtain older ages with {\it GM} than with {\it CB} and {\it BC}, but the diffences in color are very small in the spectral range of our data, only becoming noticeable at longer wavelengths (see Fig.\ 2 of BC03).

\subsection{Data: 98291 spectra from 107 CALIFA datacubes}
\label{sec:Data4SSPsTest}

The experiments described below were based on data with the same characteristics and pre-processed in the exact same way as the data for CALIFA 277. The difference with respect to the simulations presented in the first part of this paper is that we now use 107 galaxies, out of which a total of 98291 spectra corresponding to individual spaxels (or Voronoi zones) were extracted following the methodology outlined in Paper I.

\starlight\ fits of these spectra were carried out using the three bases described above. Our interest here is to identify and quantify differences in the  results obtained with different model SSPs, so we will overlook the spatial information and treat each spectrum as an individual. We now use this massive dataset to compare the results obtained with bases {\it GM}, {\it CB} and {\it BC}.


\subsection{Comparison of global properties}
\label{sec:BaseComparison}

\begin{figure*}
\centering\includegraphics[width=0.99\textwidth]{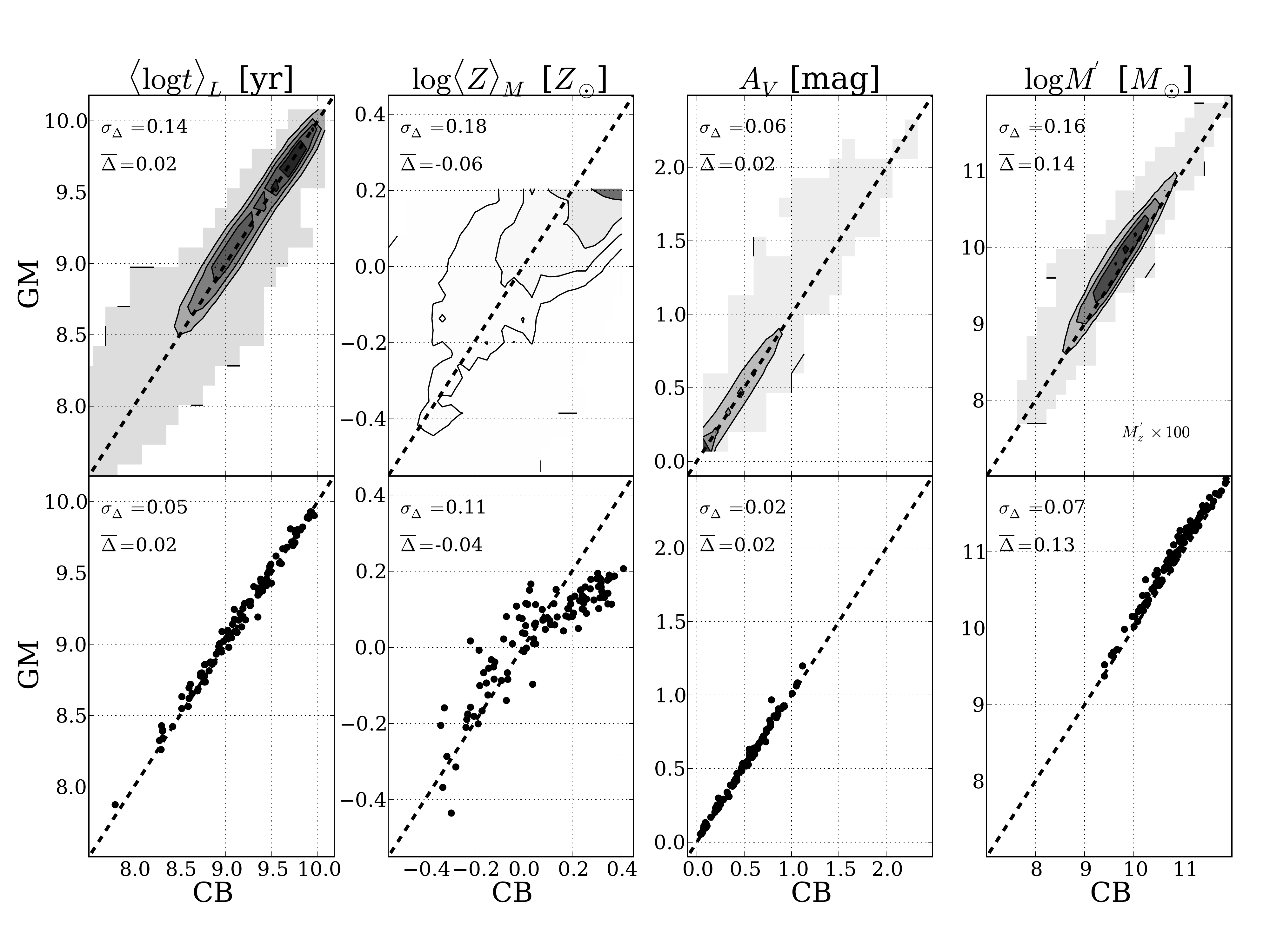}
\caption{Comparison of the global properties obtained with different  bases {\it GM} (vertical axis) and {\it CB} (horizontal). The top panels compare luminosity weighted ages ($\langle \log t \rangle_L$), mass weighted metallicities ($\log \langle Z \rangle_M$), extinction ($A_V$) and initial stellar masses ($\log M^\prime$) obtained for $\sim 10^5$ zones of 107 CALIFA galaxies, with contours drawn at every 20\% of enclosed points.  The bottom panels repeats the comparisons, but now for galaxy-wide values. A one-to-one line is drawn in all panels. Note that zone masses in the top right panel are multiplied by 100 to put them on the same scale as the bottom panel.
}
\label{fig:GlobProps_GMxCB}
\end{figure*}

\begin{figure*}
\centering\includegraphics[width=0.99\textwidth]{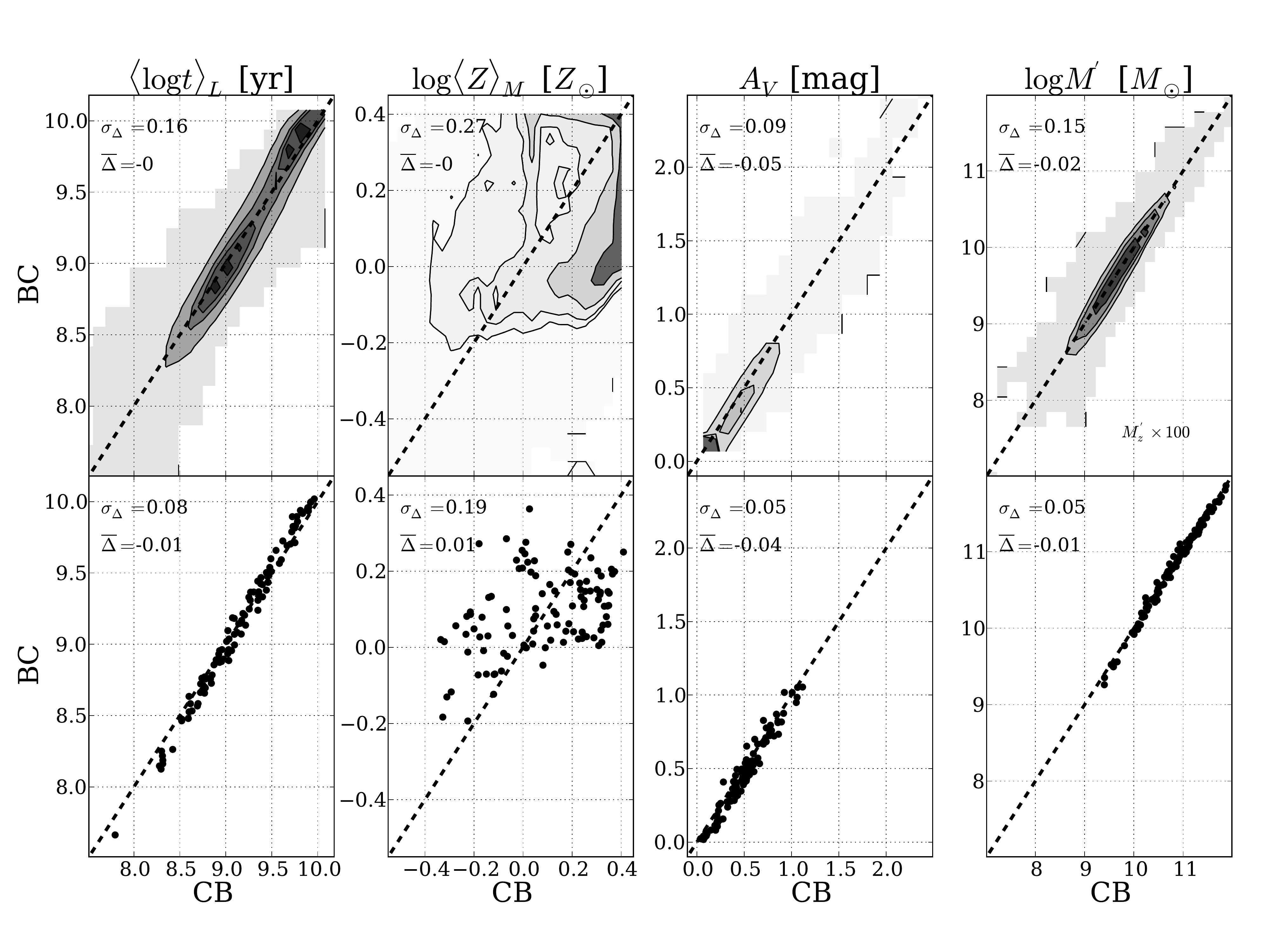}
\caption{As Fig.\ \ref{fig:GlobProps_GMxCB}, but comparing results obtained with base {\em CB} (horizontal axis) and {\em BC} (vertical). Ages, extinction and masses agree well, but BC-based mean metallicities differ significantly from both 
{\em CB} and {\em GM}-based estimates.}
\label{fig:GlobProps_BCxCB}
\end{figure*}

Fig.\ \ref{fig:GlobProps_GMxCB} compares results obtained with bases {\it CB} (horizontal axis) and  {\it GM}  (vertical). The top panels compare values of $\langle \log t \rangle_L$,  $\log \langle Z \rangle_M$, $A_V$ and $\log M^\prime$ for all 98921 individual zones, grayscale-coding by the density of points. Notice the wide ranges of values in all panels, illustrative of the diversity of properties spanned by the data. Each panel quotes the mean $\Delta$ and its standard deviation, where $\Delta = $ property({\em GM}) $-$ property({\em CB}).

{\it GM}-based initial stellar masses are higher than {\it CB} ones by 0.14 dex on-average, reflecting their different IMFs.\footnote{ Recall that $M^\prime$ refers to the initial stellar mass. The difference between {\it GM} and {\it CB} masses increases to 0.27 dex when stellar mass loss is accounted for.}
Apart from this offset, the two masses agree to within 0.16 dex. Mean ages and extinctions are also in good agreement, with dispersions of 0.14 dex in $\langle \log t \rangle_L$ and 0.06 mag in $A_V$, and insignificant offsets. Metallicities are obviously related, but with a substantial scatter due to a combination of the difference in evolutionary tracks and IMF, coarse $Z$ grids, intrinsic difficulties in estimating metallicities, and the aforementioned difference in maximum $Z$. 

To estimate the effect of this in-built difference we have carried out experiments replacing the $2.5 Z_\odot$ SSPs in {\it CB} by models interpolated to $1.5 Z_\odot$, thus matching the $Z$'s in {\it GM}. This modified {\it CB} base produces $\log \langle Z \rangle_M$ values which differ by $\sigma_\Delta \sim 0.15$ dex from the {\it GM}-based ones, slightly better than the 0.18 dex obtained with the original {\it CB}. Leaving aside caveats associated to the interpolation of SSP spectra, this result indicates that the difference in maximum $Z$ between {\it GM} and {\it CB} has a relatively modest impact on the difference between $\log \langle Z \rangle_M$ values obtained with these two bases, so that differences in evolutionary tracks and IMF are also relevant.


The bottom panels repeat the comparison of the same four properties, but now for galaxy-wide values, so there are only 107 points.\footnote{$M^\prime$ in this case is the sum of the $M^\prime_z$ values for each zone. The galaxy-wide $\langle \log t \rangle_L$ is obtained from $\sum_j \log t_j \sum_z L_{jz} / \sum_j \sum_z L_{jz} $, where $j$ is the base age index and $z$ is a zone index. $\langle Z \rangle_M$ is computed analogously. Lastly, the galaxy-wide $A_V$ is defined as the mean $A_{V,z}$ of all its zones.} These cleaner plots confirm the results from the zone-by-zone comparison. Naturally, galaxy-wide averages are much more robust than values for individual zones, so the $\sigma_\Delta$ dispersions are significantly smaller than those quoted in the top panels. $\langle Z \rangle_M$ estimates distribute around the identity line up to $Z_\odot$, where the relation bends down due to the differences in the maximum $Z$ between the bases. In these cases all one can be sure of is that metalicities are over-solar. 

Fig.\ \ref{fig:GlobProps_BCxCB} compares results for bases {\it BC} (vertical axis) with {\it CB} (horizontal). Masses, ages, and extinctions agree to within 0.15 dex, 0.16 dex and 0.09 mag respectively, with a tendency of $A_V$ to be 0.04 mag smaller in {\it BC} fits. This relatively good level of agreement does not apply to mass weighted metallicities, which differ substantially. Given that these two sets of models coincide in isochrones, IMF and evolutionary synthesis code, this disagrement must be related to the stellar libraries. Because of the several improvements in MILES over STELIB (S\'anchez-Bl\'azquez et al.\ 2006), metallicities from the {\it CB} and {\it GM} bases should be considered more reliable than those obtained with  {\it BC} (see also Koleva et al.\ 2008). This is in line with previous empirical tests by Gonz\'alez Delgado \& Cid Fernandes (2010), who showed that MILES based SSP spectra perform better than STELIB based ones in matching the metallicities of star clusters derived from spectroscopy of individual stars or CMD analysis of the same systems.

\begin{figure*}
\centering\includegraphics[width=1.0\textwidth]{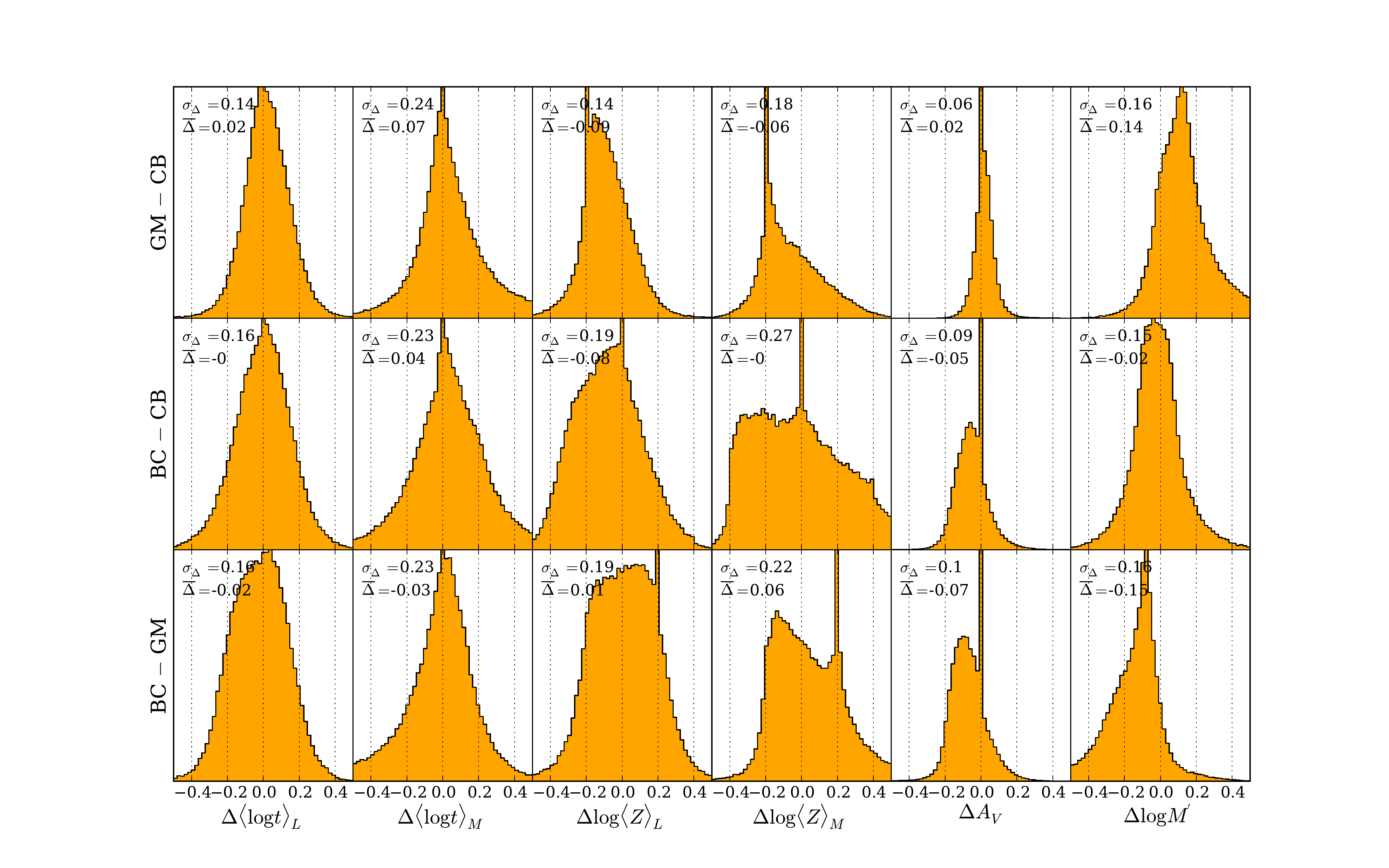}
\caption{Histograms of differences in global properties obtained with bases 
{\it GM}, {\it CB} and  {\it BC}. $\Delta$'s are defined as
{\it GM} $-$ {\it CB} in the top panels,  {\it BC} $-$ {\it CB} in the middle, and 
{\it BC} $-$ {\it GM} in the bottom ones. The horizontal scale is the same as in Fig.\ \ref{fig:DiffHists} to facilitate the comparison of data $+$ method related uncertainties with the model related ones.}
\label{fig:SSPsDiffHists}
\end{figure*}

Fig.\ \ref{fig:SSPsDiffHists} shows histograms of the pair-wise $\Delta$'s: {\em GM} $-$ {\em CB} (top), {\em BC} $-$ {\em CB} (middle), {\em GM} $-$ {\em BC} (bottom). The width of these histograms (the $\sigma_\Delta$ values listed in each panel) gives a measure of the uncertainties related to the choice of evolutionary synthesis models for the spectral fits. As usual, mass weighted properties present larger uncertainties than luminosity-weighted ones. Metallicities, for instance, have dispersions in $\Delta\log \langle Z \rangle_L$ of 0.14 and 0.19 dex ({\em GM} versus {\em CB}, and {\em BC} versus {\em CB}, respectively), smaller than the 0.18 and 0.27 dex found for $\Delta\log \langle Z \rangle_M$.

The overal conclusion of these comparisons is that, from a statistical perspective, physical properties obtained from a spectral synthesis analysis do not depend strongly on the choice of SSP models. Mean ages and extinctions retrieved with the three models investigated here all agree to within relatively small margins, and the same is true for stellar masses once in-built differences in IMFs are accounted for. Metallicities are generally harder to estimate and our experiments confirm that. Unlike  other properties mean $Z$'s are sensitive to the choice of models. Base {\em BC} gives the most discrepant results, highlighting the influence of stellar libraries.

Rounding up numbers, we find that  uncertainties related to the SSP models adopted in the analysis are of the order of 0.15 dex in  $\langle \log t \rangle_L$,  0.23 dex in  $\langle \log t \rangle_M$,  0.15 dex in $\log M^\prime$ and 0.1 mag in $A_V$. Excluding {\em BC} models, $\log \langle Z \rangle_L$ and  $\log \langle Z \rangle_M$ have uncertainties of 0.14 and 0.18 dex, respectively.

\subsubsection{Models versus data and method related uncertainties}

How do these uncertainties compare with those related to the data and synthesis method, derived in the first part of this article? 

To answer this, we focus on the OR1 and OC002 simulations described in \S\ref{sec:ErrorsAndSimulations}, as they are the ones which represent  random noise and shape-calibration errors in CALIFA. Comparing the $\sigma_\Delta$ values in Fig.\ \ref{fig:SSPsDiffHists} with those for the OR1 simulations in Fig.\ \ref{fig:DiffHists} we see that the former are roughly twice as large as the latter. In fact, the $\sigma_\Delta$ values in the top panels of Fig.\ \ref{fig:SSPsDiffHists} are closer to those obtained in the simulations where the noise was artificially amplified by a factor or 2 (OR2 runs in Table \ref{tab:Stats_SimsK0277}). Shape-related uncertainties also lead to uncertainties somewhat smaller than those found in this section, with the exception of $A_V$, whose uncertainty is 0.16 mag in the OC002 runs but 0.1 mag or less in $\Delta A_V$ histograms in Fig.\ \ref{fig:SSPsDiffHists}. A visual comparison of Figs.\ \ref{fig:DiffHists}  and \ref{fig:SSPsDiffHists} corroborates the conclusion that uncertainties related to differences among SSP models are larger than those related to data and method. 

These comparisons, however, are not entirely fair, since model related uncertainties were evaluated from $\sim 10^5$ zones of 107 galaxies, while data $+$ method related uncertainty estimates were based on 1638 zones of a single galaxy, perturbed 10 times in the simulations. Redoing Fig.\ \ref{fig:SSPsDiffHists} for CALIFA 277 alone we still find $\sigma_\Delta$ values higher than those found in \S\ref{sec:SimulationsResults}, but by a smaller margin. The important message, therefore, is not so much that variance among current SSP spectral models leads to larger uncertainties than those related to data $+$ method, but that both kinds of uncertainties are of the same order.

\subsection{Quality of spectral fits}
\label{sec:SpectralResiduals}

\begin{figure}
\centering\includegraphics[width=0.5\textwidth]{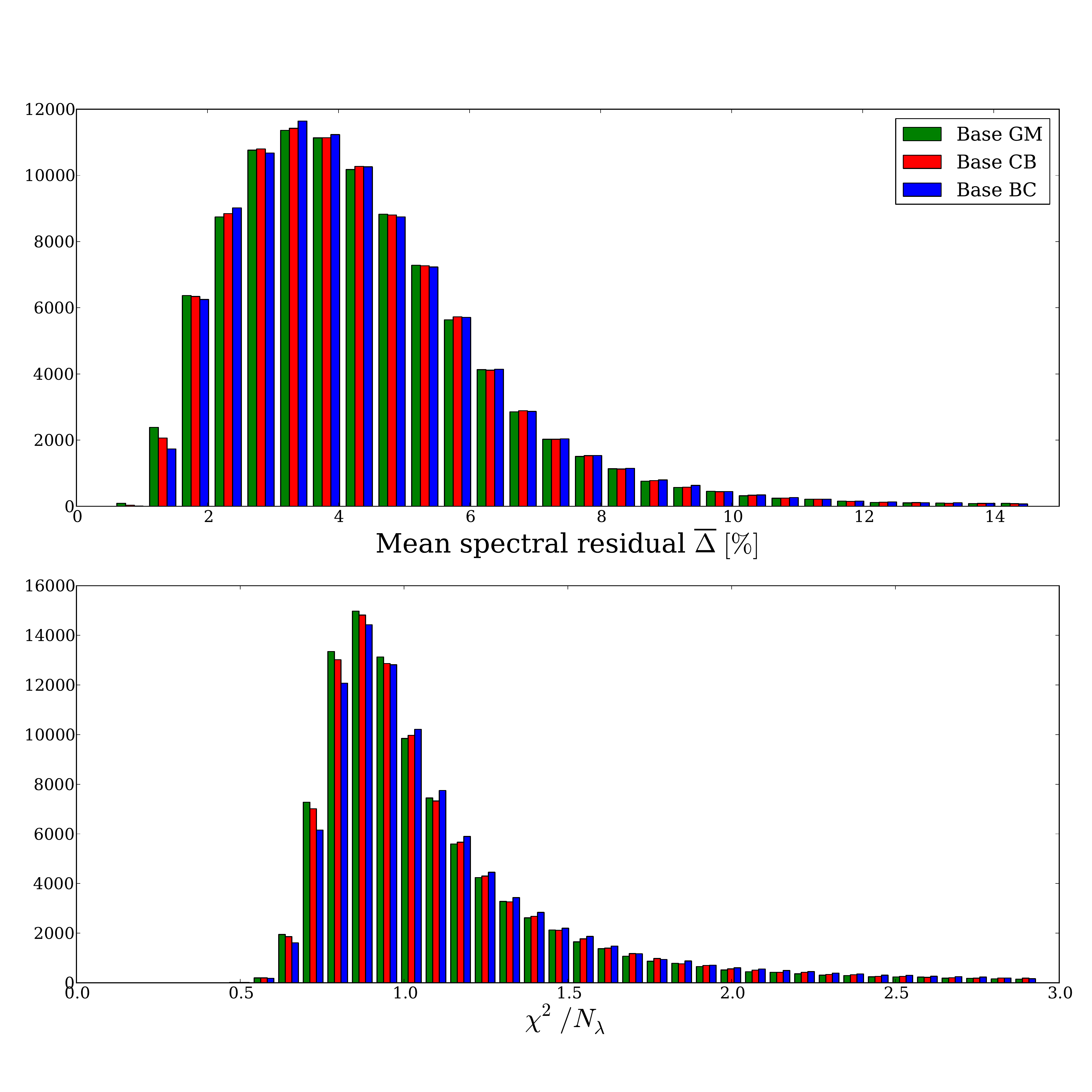}
\caption{{\it Top:} Distribution of the  mean percentual spectral deviation 
for \starlight\ fits with the {\it GM} (bars in green),  {\it CB} (red) and  {\it BC} (blue) bases. {\it Bottom:} As above, but for the distribution of $\chi^2$ per fitted flux. In both cases the distributions are essentially identical, reflecting the fact that the three bases provide equally good spectral fits.}
\label{fig:adevHist}
\end{figure}

\begin{figure}
\centering\includegraphics[width=0.5\textwidth]{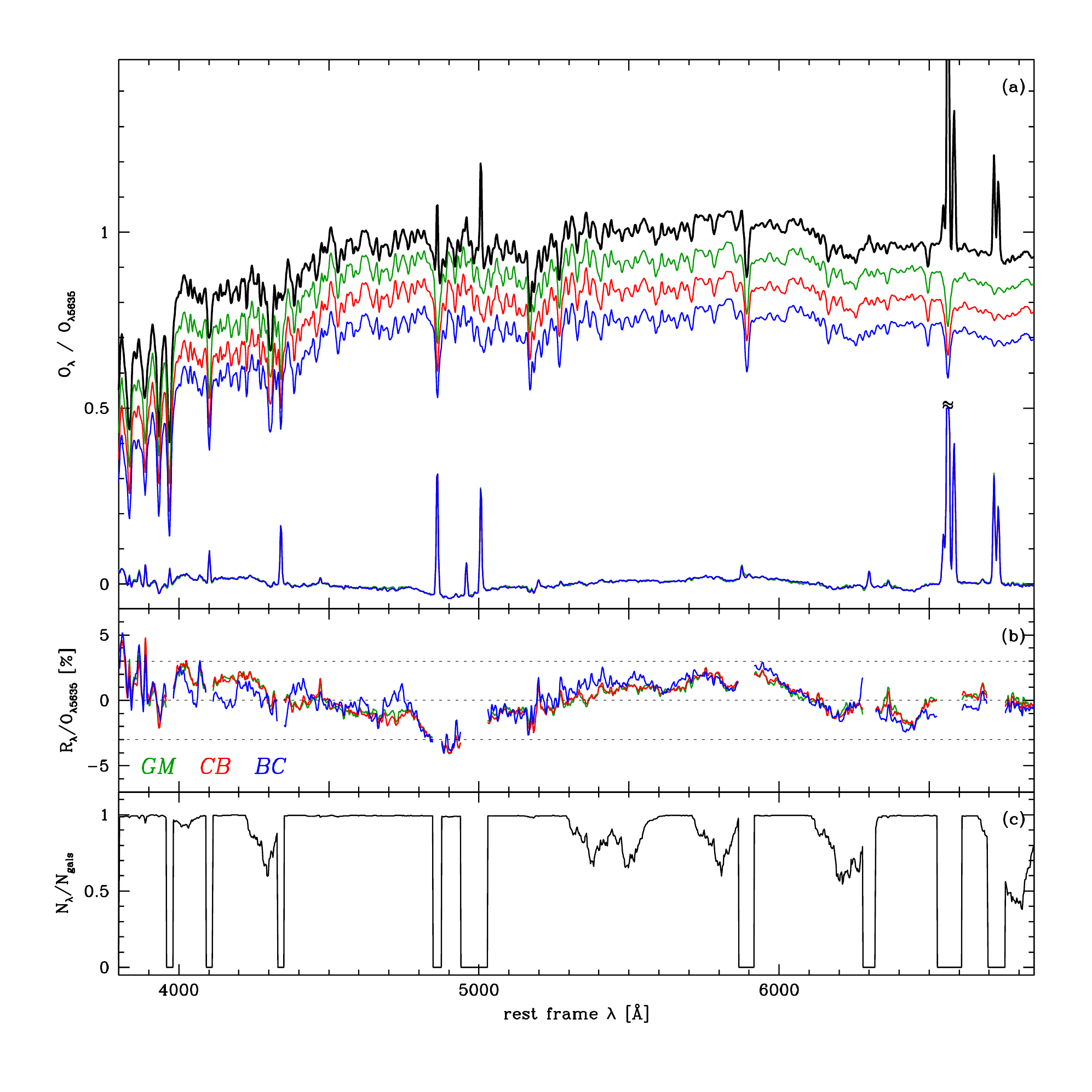}
\caption{{\it (a)} The black line shows the mean (``stacked'') normalized spectrum of 98291 zones from 107 CALIFA galaxies, obtained by averaging $O_\lambda / O_{\lambda5635}$. Green, red and blue lines (offset for clarity) show the mean synthetic spectra for bases {\it GM}, {\it CB} and {\it BC}, respectively. The corresponding $(O_\lambda - M_\lambda) / O_{\lambda5635}$ residual spectra are shown with the same colors, but they are hardly distinguishable. {\it (b)} Zoom of the spectral residuals in the top panel, with emission lines removed for clarity. Dashed lines mark residuals of $-3$, 0 and $+3$\%. {\it (c)} Fraction of the 98291 zones contributing to the spectral statistics in the midde panel.}
\label{fig:SpecStats98291zones}
\end{figure}

In principle one could use the quality of the spectral fits as a criterion to favor one base over another. In practice, however, all bases studied here provide $\sim$ equally good fits to the data. This is illustrated in  Fig.\ \ref{fig:adevHist}, which shows the distributions of two figures of merit. Green, red and blue correspond to bases  {\it GM}, {\it CB} and {\it BC}, respectively. The histograms are nearly identical. Considering all 98291 fits, the mean (median) values of the relative spectral deviation (eq.\ 6 of Paper I) are $\overline{\Delta} = 4.33$ (3.92), 4.34 (3.93) and 4.37 (3.94) \% for bases {\it GM}, {\it CB} and {\it BC} respectively, while the corresponding mean (median) $\chi^2$ per fitted flux are 1.14 (0.96), 1.15 (0.97) and 1.18 (0.99). Formally, this puts base {\it GM} in $1^{st}$ place, with {\it CB} in $2^{nd}$ and {\it BC} in $3^{rd}$, but these statistics reveal that it is clearly not possible to favour one or another model in terms of fit quality. Because of their more complete spectral libraries, we favor bases {\it GM} and {\it CB} over {\it BC}.

Fig.\ \ref{fig:SpecStats98291zones} provides a more direct visualization of the quality of the spectral fits. The top panel shows (in black) the average of all zone spectra  (for examples of fits to individual zones see Gonz\'alez Delgado et al., in prep.). Each spectrum is first divided by the median flux in the 5590--5680 \AA\ normalization window, such that all zones weight equally in the average. The average synthetic spectra are shown in green, red and blue for fits with bases  {\it GM}, {\it CB} and {\it BC}, respectively, vertically offset for clarity. The corresponding residual spectra are shown in the bottom of the same panel, but can hardly be distinguished on this scale. This similarity persists even after zooming in by a full order of magnitude in the flux scale, as shown in Fig.\ \ref{fig:SpecStats98291zones}b. Gaps in this zoomed plot correspond to our generic emission line mask, ie., $\lambda$-windows which were ignored in the \starlight\ fits. Notice that, because of bad pixel flags, augmented by windows around telluric features (Paper I), the number of specra involved in these averaged residuals varies for each $\lambda$, as shown in the bottom panel.

\subsection{Spectral residuals and comparison with SDSS}
\label{sec:Comparison2SDSS}

Clearly, the amplitude of the average spectral residuals is very small. Whereas the typical relative deviation for fits of individual zones is $\overline{\Delta} \sim 4$\% (Fig.\ \ref{fig:adevHist}), for the average spectra in Fig.\ \ref{fig:SpecStats98291zones} it is a tiny $\overline{\Delta} = 1$\%. This 4-fold reduction indicates that random noise is responsible for a large share of the residuals in spectral fits of individual zones. Yet, if the residuals were truly random, this reduction should be by a factor of the order of $\sqrt{98291} \sim 300$, not just 4. Residuals may be small, but they are definitely real.

The systematics of the spectral residuals are visible in 
Fig.\ \ref{fig:SpecStats98291zones}b. These deviations must be related to issues in the method, data and/or models. 

The narrow positive peaks in the average residual spectrum are due to our fitting method. They are in fact emission lines not included in our generic emission line mask. [NeIII]$\lambda$3869, H8$\lambda$3889, [NI]$\lambda$5199 and [OI]$\lambda$6364 are the clearest ones in Fig.\ \ref{fig:SpecStats98291zones}b, but weaker lines such as H9$\lambda$3835 and HeI$\lambda$6678 are identified upon closer inspection.  The conservative clipping parameters adopted in the fits (see Paper I) prevent \starlight\ from authomatically clipping these lines completely. Other \starlight-based studies circumvent this problem by building emission line masks taylored for each spectrum individually (Mateus et al.\ 2006; Asari et al.\ 2007), a refinement which is yet to be implemented in our CALIFA pipelines.

\begin{figure*}
\centering\includegraphics[width=1.0\textwidth]{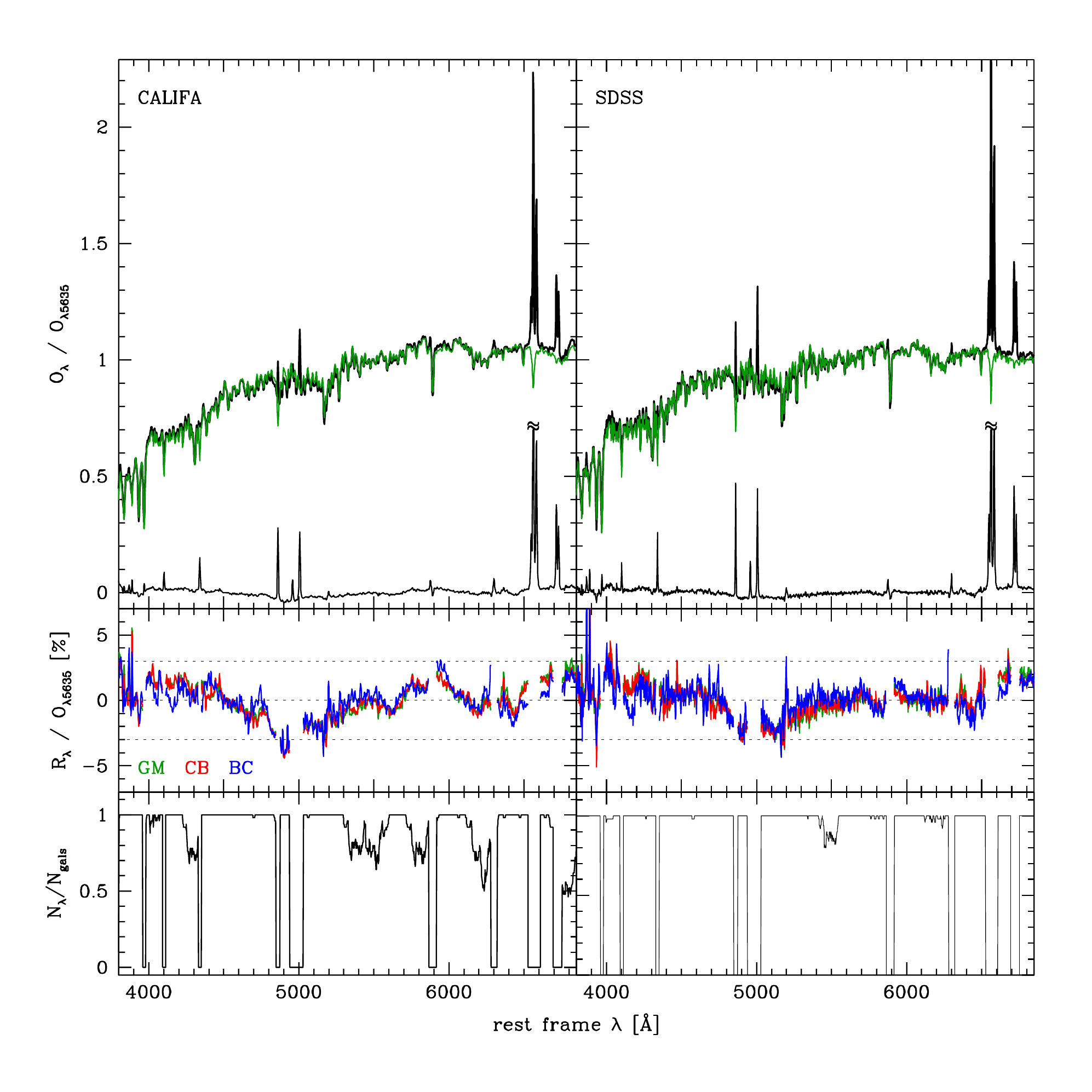}
\caption{As Fig.\ \ref{fig:SpecStats98291zones}, but for a subset of 50 CALIFA galaxies for which SDSS spectra are available. Unlike in Fig.\ \ref{fig:SpecStats98291zones}, however, the top panels show only results for {\it GM} based spectral fits. Left panels show the spectral statistics for the nuclear extractions of CALIFA datacubes, whereas those on the right show results obtained from \starlight\ fits to the SDSS spectra.}
\label{fig:SpecStats50gals}
\end{figure*}

The most noticeable features in Fig.\ \ref{fig:SpecStats98291zones}b, however, are not the emission lines, but the broad $O_\lambda < M_\lambda$ trough around H$\beta$ and the $O_\lambda > M_\lambda$ bumps in the 5800--6000 \AA\ region. In order to understand whether these features are particular to CALIFA data we have compiled SDSS spectra for 50 galaxies in common with the current sample and analysed them using the same masks, wavelength range, sampling, and \starlight\ configuration as used for CALIFA. These spectra were then averaged as in Fig.\ \ref{fig:SpecStats98291zones}, and compared to the spectral statistics obtained from the nuclear CALIFA spectra of the same 50 galaxies. Fig.\ \ref{fig:SpecStats50gals} shows the results of this comparison. The figure  follows the same layout as Fig.\ \ref{fig:SpecStats98291zones}, with spectral statistics for the 50 CALIFA nuclei and the corresponding SDSS spectra plotted on the left and right panels, respectively. The spikier appearance of the SDSS residuals is due to its higher spectral resolution (3 versus 6 \AA\ FWHM), but this is irrelevant for the broad residual features discussed above.

The humps around the $\sim 5800$ \AA\ region of CALIFA residuals are clearly absent in the SDSS. These features are seen as shallow slanted bands in Fig.\ 16 of Husseman et al.\ (2013), where residuals for 100 galaxies are sorted by redshift. We have tracked the origin of this feature to an imperfect removal of telluric NaI  bands around 5900 \AA\ in the observed frame. In analogy with what happened with the SDSS data releases, as CALIFA data is used, imperfections like this will be identified and, when possible, corrected for in future updates of the data base. 

The trough around H$\beta$, however, is present in both SDSS and CALIFA spectra. In fact, this problem has been noticed before (Cid Fernandes 2006; Panter et al.\ 2007; Walcher et al.\ 2009), but not satisfactorely explained to date. The fact that this feature appears in both data sets suggests that the models overpredict the continuum in a $\sim 200$ \AA\ wide region centered around H$\beta$. On the other hand, the fact that the problem is present for both MILES and STELIB based models suggest that its origin is more likely related to data calibration issues. This latter view is supported by studies like Ferr\'e-Mateu et al.\ (2012), where \starlight\ fits of William Herschel Telescope long-slit spectra of massive compact galaxies with the Vazdekis et al.\ (2010) models (included in our {\it GM} base) do not reveal the H$\beta$ trough seen in Fig.\ \ref{fig:SpecStats98291zones}b. While a full investigation of this issue is beyond the scope of this paper, we note that the trough tends to be more pronounced in younger systems, as previously reported by Cid Fernandes (2006, see his Figure 5). 

The negative residual seen in the region of Mg lines near 5175 \AA\ is almost as strong as the H$\beta$ trough. This deficit is due to differences in the abundances of $\alpha$ elements between galaxies (particular the central regions of massive ones) and stars in the stellar libraries. The fact that this residual is stronger in Fig.\ \ref{fig:SpecStats50gals} (nuclear spectra)  than in Fig.\ \ref{fig:SpecStats98291zones} (all zones) is likely due to $\alpha$/Fe gradients. Walcher et al.\ (2009) report steps towards incorporating $\alpha$/Fe in spectral synthesis analysis.

\subsection{Public distribution of base GM spectra}

Given the good performance of the {\it GM} base in the tests discussed in this section, we make these spectra available to the community in www.iaa.es/$\sim$rosa/AYA2010/AYA2010. We stress that these models were produced by Gonz\'alez Delgado et al.\ (2005) and Vazdekis et al.\ (2010), to whom all credit should be given. Our distribution just puts these two sets of models together.

\section{Summary}
\label{sec:Conclusions}

We have performed an extensive study of the uncertainties in the stellar population properties derived from \starlight\ fits of optical integral field spectra from the CALIFA survey. In the first part of this investigation,  Monte Carlo simulations were used to explore the effects of random noise, continuum shape calibration uncertainties and degeneracies in the spectral synthesis. The simulations are based on the data and \starlight\ fits of 1638 zone spectra of the Sb galaxy CALIFA 277 (NGC 2916). Uncertainties in mean age, metallicity, extincion, stellar mass and SFHs were evaluated both at the level individual zones (treating the datacube as a collection of unrelated spectra) and exploring the statistical power of IFS  by means of radial profiles and other spatial averages. Second, we explore the variations in physical properties due to the use of different sets of models for SSP spectra. We compare physical properties and spectral residuals obtained from fits with three different SSP bases: (a) a combination of Gonzalez Delgado et al.\ (2005) and Vazdekis et al.\ (2010) models ({\it GM}); (b) preliminary Charlot \& Bruzual models using the {\sc granada} and MILES libraries ({\it CB}); (c) the standard Bruzual \& Charlot (2003) models with STELIB ({\it BC}). This comparison employs $\sim 10^5$ spectra from 107 CALIFA galaxies.

Our main results can be summarized as follows:

\begin{enumerate}

\item The level of noise in CALIFA spectra leads to one-$\sigma$ uncertainties 
in mean ages and metallicities of $\sim 0.08$ dex when weighting in luminosity and $\sim 0.15$ dex weighting by mass. Stellar masses are uncertain at the $\sim 0.08$ dex level, while $\sigma(A_V)$ is about 0.06 mag.

\item  Shape-related calibration uncertainties of 0.05 mag in $g - r$ produce uncertainties of the same order of those induced by random noise, except for $A_V$, which becomes uncertain at the level of $\sigma \sim 0.16$ mag and biased by $+0.05$ mag. 

\item Time resolved SFHs are higher order products of the spectral synthesis, inevitably subjected to larger uncertainties than global properties. For instance, we find that, for individual zones,  stellar masses formed in broad age ranges have uncertainties about 3 times larger than those of the mass formed over all times. While SFHs of individual zones do carry useful information, a statistical approach is more advisable.

\item Spectral synthesis results are best used in comparative studies involving large samples, and IFS data fit this requirement perfectly. Spatial averaging reduces uncertainties while preserving enough information on the history and structure of stellar populations. Radial profiles of mean ages, metallicities, mass density and extincion are found to be very robust due to the large number statistics of datacubes. For instance, uncertainties in the data and limitations of the synthesis have a negligible impact upon the negative age and metallicity gradients detected in CALIFA 277. Similarly, SFHs averaged over different spatial regions are much more stable than those obtained for individual zones.

\item Mean ages, extinction and stellar masses obtained with {\it GM}, {\it CB} and {\it BC} models are reasonably consistent with one another, with one-$\sigma$ differences in $\langle \log t \rangle_L$, $\langle \log t \rangle_M$, $\log M^\prime$ and  $A_V$ of $\sim 0.15$, 0.24, 0.15 dex and 0.08 mag, respectively. {\it GM} and {\it CB} produce luminosity (mass) weighted metallicities consistent to within 0.14 (0.18) dex rms, while {\it BC}-based metallicities hardly correlate with the others.

The variations induced by the use of different SSP models are larger than uncertainties due to data and method by roughly a factor of 2, highlighting the important role of this ingredient in the whole analysis.

\item The quality of the spectral fits is very nearly the same whichever set of SSP models is used. Spectral residuals are of the order of 4\% for individual zones and 1\% when averaged over all zones and galaxies. 
The strongest feature in the residual spectra is a broad but shallow ($\sim 3$\%) trough around H$\beta$, also present in SDSS spectra, and its origin remains unidentified. Other systematic features in the residual spectra were tracked to incomplete masking of weak emission lines, imperfect removal of telluric features and the lack of $\alpha$ enhanced models in the base. 

\end{enumerate}

These results, along with the tables and figures in this paper, fulfill our goal of assessing the uncertainties on the several products of \starlight\ applied to CALIFA datacubes presented in Paper I. The combination of these spatial and temporal diagnostics of stellar populations with the diversity of galaxy types in the CALIFA survey offers new ways to study galaxy evolution, which shall be explored in other publications by our collaboration, and this paper provides quantitative and qualitative guidelines on how to interpret them appropriately.

\begin{acknowledgements} 

CALIFA is the first legacy survey being performed at Calar Alto. The CALIFA collaboration would like to thank the IAA-CSIC and MPIA-MPG as major partners of the observatory, and CAHA itself, for the unique access to telescope time and support in manpower and infrastructures.  We also thank the CAHA staff for the dedication to this project.

RCF thanks the hospitality of the IAA and the support of CAPES and CNPq. ALA acknowledges support from INCT-A, Brazil. BH gratefully acknowledges the support by the DFG via grant Wi 1369/29-1.  Support from the Spanish Ministerio de Economia y Competitividad, through projects AYA2010-15081 (PI RGD), AYA2010-22111-C03-03 and AYA2010-10904E (SFS), AYA2010-21322-C03-02 (PSB) and the Ram\'on y Cajal Program (SFS, PSB and JFB), is warmly acknowledged. We also thank the Viabilidad , Dise\~no , Acceso y Mejora funding program, ICTS-2009-10, for funding the data acquisition of this project. 

Funding for the SDSS and SDSS-II has been provided by the Alfred P. Sloan Foundation, the Participating Institutions, the National Science Foundation, the U.S. Department of Energy, the National Aeronautics and Space Administration, the Japanese Monbukagakusho, the Max Planck Society, and the Higher Education Funding Council for England. The SDSS Web Site is http://www.sdss.org/.

The SDSS is managed by the Astrophysical Research Consortium for the Participating Institutions. The Participating Institutions are the American Museum of Natural History, Astrophysical Institute Potsdam, University of Basel, University of Cambridge, Case Western Reserve University, University of Chicago, Drexel University, Fermilab, the Institute for Advanced Study, the Japan Participation Group, Johns Hopkins University, the Joint Institute for Nuclear Astrophysics, the Kavli Institute for Particle Astrophysics and Cosmology, the Korean Scientist Group, the Chinese Academy of Sciences (LAMOST), Los Alamos National Laboratory, the Max-Planck-Institute for Astronomy (MPIA), the Max-Planck-Institute for Astrophysics (MPA), New Mexico State University, Ohio State University, University of Pittsburgh, University of Portsmouth, Princeton University, the United States Naval Observatory, and the University of Washington.

\end{acknowledgements}

\end{document}